# Huge hole injection in tungsten dichalcogenide heterostructures without electric gating: a DFT study

**Dawid Ciszewski[1*] and Wojciech Grochala[1*]**

[1] Centre of New Technologies University of Warsaw, Żwirki i Wigury 93, 02-089 Warsaw, Poland

*Authors to whom any correspondence should be addressed.

**E-mail:** d.ciszewski@cent.uw.edu.pl, w.grochala@cent.uw.edu.pl

## Abstract

Van der Waals heterostructures based on transition metal dichalcogenides (TMDs) provide a versatile platform for tailoring electronic properties through interlayer charge transfer (CT). Precise control of CT is essential because it directly determines the electronic structure and carrier concentration in atomically thin materials. Recently, the concept of a chemical capacitor has been proposed as a route to achieving exceptionally high carrier densities through CT across insulating separator layers. Here, we extend this concept to van der Waals heterostructures by investigating TMD | hBN | OX (oxidizer) systems using density functional theory (DFT). Following the screening of candidate TMDs and electron acceptors, $XeF_2$ and $KrF_2$ were identified as suitable acceptors exhibiting type-III (broken-gap) band alignment with $WS_2$ and $WSe_2$, respectively. Periodic DFT calculations of large supercells reveal CT corresponding to hole concentrations of up to 0.23 $h^+$ and 0.35 $h^+$ per W atom in $WS_2$ | hBN | $XeF_2$ and $WSe_2$ | hBN | $KrF_2$ heterostructures, respectively. The resulting charge redistribution demonstrates that noble-gas fluorides provide an efficient route for non-contact engineering of carrier density in TMD heterostructures, offering a new strategy for tuning correlated electronic phases in two-dimensional materials.

## 1. Introduction

The ability to precisely control carrier density has become one of the central paradigms of modern condensed matter physics, providing a versatile route towards tailoring the electronic structure and inducing a broad spectrum of electronic phenomena in low-dimensional materials. Several approaches have been developed to achieve electron and hole doping in two-dimensional materials, including electrostatic gating, substitutional doping, molecular adsorption and interfacial charge transfer (CT)[1]-[3]. While electrostatic gating enables reversible tuning of carrier concentration, it requires the continuous application of an external electric field. Conventional chemical doping, on the other hand, frequently introduces structural disorder, impurity scattering, degradation of carrier mobility or limited long-term stability. Consequently, contactless approaches capable of inducing stable and controllable carrier concentrations without external bias or structural modification of the active material add a new solution to the set of doping methods.

One such solution is the recently proposed concept of a chemical capacitor (CC)[3], in which spontaneous CT occurs between an electron donor (reducer RED) and an electron acceptor (oxidizer OX) separated by an atomically thin dielectric separator (SEP), forming a RED | SEP | OX nanostructure. Unlike electrostatic gating or conventional chemical doping, the active material is neither continuously driven by an externally applied electric field nor permanently chemically modified. Instead, carrier injection is driven solely by the chemical potential difference between the donor and acceptor layers. Moreover, the amount of transferred charge can, in principle, be controlled by the electronic properties of the constituent materials as well as by the thickness of the dielectric separator.

Theoretical studies performed so far have predicted substantial electron and hole doping in several CC architectures, including metal | halide | halogen[3] and hydride systems[4] bearing very reactive components. However, despite their attractive calculated properties, these computationally convenient model structures remain difficult to realize experimentally owing to challenges associated with material compatibility, fabrication and mechanical integration of the individual nanolayers. Therefore, identifying experimentally realistic material platforms capable of implementing the CC concept constitutes the next essential step towards its practical realization.

Importantly, van der Waals heterostructures naturally satisfy many of the structural and electronic requirements of the CC architecture[1]-[3][5]. In particular, hBN has become the dielectric material of choice in two-dimensional heterostructures owing to its large band gap, chemical inertness and atomically flat

interfaces[6]-[12]. Numerous studies have demonstrated that hBN effectively suppresses direct electronic hybridization while preserving electrostatic coupling between neighbouring layers[1][12]-[14]. Furthermore, both the magnitude and the dynamics of interlayer coupling can be precisely tuned by varying the number of intermediate hBN layers[14][15]. This operating principle closely resembles that proposed for the chemical capacitor and therefore suggests that van der Waals heterostructures may provide the first realistic experimental platform for its implementation[9][16].

In this work, we investigate experimentally feasible TMD | hBN | OX heterostructures as candidate implementations of the CC architecture. A computational screening of a broad range of TMD/acceptor combinations was first performed by analysing their electronic DOS and band alignment with respect to a common energy reference in order to identify systems expected to exhibit the strongest spontaneous CT. Noble-gas fluorides were considered as molecular electron acceptors owing to their high electron affinity. Subsequently, selected $WS_2$ | hBN | $XeF_2$ and $WSe_2$ | hBN | $KrF_2$ heterostructures were investigated using density functional theory (DFT) in periodic supercells to quantify interfacial CT and characterize the resulting modification of their crystal and electronic structure.

## 2. CT in TMD van der Waals Heterostructures reported so far

Transition metal dichalcogenides constitute one of the most versatile material platforms for engineering electronic properties in atomically thin semiconductors, thus enabling applications ranging from electronic and optoelectronic devices to sensors, catalysts, flexible electronics, spintronics and energy storage[1][17]-[25]. Owing to their tunable band structure, strong Coulomb interactions and compatibility with van der Waals integration, engineering of carrier density in TMDs has become a powerful strategy for tailoring optical, electronic and correlated phenomena, including excitonic effects, valley polarization, charge-density waves, superconductivity and catalytic activity[3][18][21][26][27]. The broad range of physical phenomena observed in TMD monolayers has stimulated intensive efforts toward precise control of their carrier density and interlayer charge transfer. As a result, numerous approaches have been developed[1][17][28] to control charge transfer and carrier concentration in TMD monolayers depending on the required application and desired doping level[13][29].

The most popular methods of inducing charge injection in TMDs include charge transfer via electrostatic gating, providing direct control over carrier concentration through an external electric field, often using electric double layer transistors[30]-[32]. Permanent carrier injection can be achieved through physisorption, where organic donor or acceptor molecules and ions transfer charge to the semiconductor while remaining bound predominantly through van der Waals interactions, leading to non-covalent functionalization of the surface and molecular adsorption[28][33]-[40]. In comparison, chemisorption and substitutional doping rely on direct chemical bonding or replacement of lattice atoms, enabling stronger and generally more stable doping but frequently accompanied by defect formation, impurity scattering or modifications of the crystal lattice through covalent functionalization[29][41]-[48]. Many such impurity heteroatoms can be injected during the growth process of TMDs, leading to disruption of the lattice structure, which may be advantageous for specific applications, such as catalysts and energy generation, where substitutional doping adds additional active sites and increases electrochemical activity of TMDs[49][50]. An extreme case of introducing foreign atoms into TMDs is alloying, where a sufficiently high concentration of substitutional atoms gives rise to a new material with intermediate properties and a modified crystal lattice[18][51]. Such alloying can also lead to degenerate doping, involving loss of the semiconductor characteristic of the TMD layers and near-metallic behaviors[24]. Other doping methods include ion implantation, photoinduced doping or intercalation, where in the latter foreign atoms or ions are inserted between layers in TMD heterostructures[26][57]. Representative carrier densities and charge-transfer values reported for these doping methods are summarized in Table 1.

As can be noted in the studies cited in Table 1, various doping methods result in CT ranging from 0.0004 to 0.44 $h^+$, and between 0.0019 and 0.07 $e^-$ per TM atom. Most approaches deliver small to moderate CT with the exception of electric gating to $WSe_2$[30] and $MoS_2$[31] which result in the upper abovementioned limits of CT. This enables tuning of the Fermi level over a broad range, leading to modifications of transport and optical properties as well as the emergence of metallic, magnetic and superconducting states. An alternative route for modifying the electronic properties of TMDs is via use of external hydrostatic pressure[58]-[61]. Compression decreases the interlayer spacing and enhances orbital overlap, resulting in progressive band-gap narrowing, increased interlayer hybridization and, above the critical pressure, semiconductor-semimetal or semiconductor-metal transitions. In several group VI TMDs these electronic transitions are accompanied by structural phase transformations and the appearance of superconducting phases. Unlike electrostatic or chemical doping, pressure preserves the total number of electrons and modifies carrier populations through induced band overlap rather than net interfacial charge transfer. Pressure therefore provides an intrinsic method for tuning the electronic structure without changing the chemical composition of the material, similarly to the principle of the CC setup. However, just like in the case of the electric gating where permanent external voltage is needed to generate the desired effect, here the same applies to the external pressure.

Building on these advances in doping and new synthesis methods[3], increasing attention has been devoted to charge transfer across van der Waals interfaces, where the electronic structure can be controlled through band alignment, dielectric screening, and interlayer coupling[5][62]-[66]. Compared with direct molecular adsorption or substitutional doping, spatial separation of the active TMD layer from the charge reservoir offers the possibility of modifying carrier concentration while reducing direct perturbation of the semiconductor lattice. This strategy provides the basis for the systems investigated in the present work[3][16].

**Table 1. Comparison of representative carrier-doping approaches reported for TMDs, including the dopant species, substrate, charge transferred, and equivalent carrier density. Wherever the value of charge transferred has not been provided we have provisionally computed its upper limit by assuming the presence of a TMD monolayer and direct proportionality of charge and carrier density with respect to the first entry in the table.**

| Doping method | Dopant | TMD | Charge/TM atom (\|e\|) | Carrier density | Reference |
|---|---|---|---|---|---|
| Electric gating | — | $WSe_2$ | 0.44 $h^+$ | $1.9 \times 10^{15}$ $cm^{-2}$ | [30] |
| | | $MoS_2$ | 0.02-0.07 $e^-$ | 1-3 $\times 10^{14}$ $cm^{-2}$ | [31] |
| | | $WSe_2$ | 0.03 $h^+$ | $1.5 \times 10^{14}$ $cm^{-2}$ | [32] |
| Molecular physisorption | Me-OED | $MoS_2$ | 0.059 $e^-$ | $1.1 \times 10^{14}$ $cm^{-2}$ | [34] |
| | BV | $MoSe_2$ | ~0.02 $e^-$ | $1.2 \times 10^{13}$ $cm^{-2}$ | [36] |
| | TTF | $MoS_2$ | 0.0047 $e^-$ | $1.4 \times 10^{13}$ $cm^{-2}$ | [36] |
| | TTF (S side) | MoSSe | 0.005 $e^-$ | $5.4 \times 10^{12}$ $cm^{-2}$ | [37] |
| | TTF (Se side) | MoSSe | 0.002 $e^-$ | $1.8 \times 10^{12}$ $cm^{-2}$ | [37] |
| | DMPD | $MoS_2$ | 0.0043 $e^-$ | $1.3 \times 10^{13}$ $cm^{-2}$ | [38] |
| | CuPc | $MoS_2$ | 0.0020 $e^-$ | $6.0 \times 10^{12}$ $cm^{-2}$ | [39] |
| | TiOPc | $MoS_2$ | 0.0019 $e^-$ | $6.0 \times 10^{12}$ $cm^{-2}$ | [39] |
| | $C_{60}F_{48}$ | $WSe_2$ | ~0.02 $h^+$ | — | [40] |
| | TCNQ | $MoS_2$ | 0.0010 $h^+$ | $1.1 \times 10^{12}$ $cm^{-2}$ | [38] |
| | TCNE | $MoS_2$ | 0.0007 $h^+$ | $2.0 \times 10^{12}$ $cm^{-2}$ | [38] |
| Chemisorption | $NO_2/NO_x$ | $WSe_2$ | 0.001 $h^+$ | $1.6 \times 10^{19}$ $cm^{-3}$ (bulk) | [42] |
| | $XeF_2$ | $WSe_2$ | 0.0002 $h^+$ | $2.8 \times 10^{18}$ $cm^{-3}$ (bulk) | [29] |
| Alloying/ Substitutional doping | W | $MoSe_2$ | 0.0004 $h^+$ | $4.0 \times 10^{11}$ $cm^{-2}$ | [43] |
| | Nb | $MoS_2$ | 0.0004 $h^+$ | $4.8 \times 10^{11}$ $cm^{-2}$ | [52] |
| | K | $WSe_2$ | 0.002 $e^-$ | $2.5 \times 10^{12}$ $cm^{-2}$ | [53] |
| | Nb | $MoS_2$ | 0.21 $h^+$ | $4.0 \times 10^{14}$ $cm^{-2}$ | [54] |
| | Re | $MoS_2$ | 0.002 $e^-$ | $2.1 \times 10^{12}$ $cm^{-2}$ | [55] |
| | Fe | $MoS_2$ | 0.002 $e^-$ | $2.4 \times 10^{12}$ $cm^{-2}$ | [56] |
| Chemical capacitor | $XeF_2$ | $WS_2$ | 0.23 $h^+$ | $2.6 \times 10^{14}$ $cm^{-2}$ | This work |
| | $KrF_2$ | $WSe_2$ | 0.35 $h^+$ | $3.7 \times 10^{14}$ $cm^{-2}$ | This work |

## 3. Methodology

The density functional theory (DFT) calculations performed to study TMD | hBN | OX systems described in this work were carried out using VASP 5.4.4 (Vienna Ab initio Simulation Package)[67]-[69]. The projector-augmented wave (PAW) method[70]-[72] was used to describe the interaction between valence electrons and ionic cores. The exchange–correlation energy was treated using the SCAN meta-GGA functional[73], supplemented by the non-local rVV10 van der Waals correction[74] to provide a balanced description of covalent bonding and long-range van der Waals interactions in layered heterostructures. Because the

heterostructures are asymmetric and exhibit a finite out-of-plane dipole moment during calculations, a dipole correction was applied along the out-of-plane direction[75][76]. A plane-wave kinetic energy cutoff of 600 eV was used for all systems providing good energy convergence. The self-consistent-field convergence threshold was set to $10^{-7}$ eV, and the geometry optimizations were considered converged when the norms of all the atomic forces were below 0.02 eV/Å for very large supercells and 0.001 eV/Å in the remaining cases. Γ-centered 6×6×1 and 24×24×1 k-point meshes were used for structural relaxation and PDOS calculations, respectively.

The $WS_2$ | hBN | $XeF_2$ system, containing 119 atoms, and the $WSe_2$ | hBN | $KrF_2$ system, containing 390 atoms, were modelled using periodic *P*1 supercells with a vacuum region of 10 Å along the out-of-plane direction to eliminate spurious interactions between periodic images. The unit cells of the TMDs and hBN were matched by minimizing the strain imposed on the 2D layers using CellMatch[77]. The final number of atoms in each supercell is therefore a consequence of the imposed threshold on the relative geometric strain, which was kept below 1%. The difference between the unit cell vectors for all components of the heterostructure implies that the stoichiometry of both studied systems is somewhat different; thus, the $WS_2$ | hBN | $XeF_2$ system corresponds to 16 $WS_2$ : 25 hBN : 7 $XeF_2$ stoichiometry (the OX:RED ratio is 0.4375), while the $WSe_2$ | hBN | $KrF_2$ system corresponds to 49 $WSe_2$ : 81 hBN : 27 $KrF_2$ (the OX:RED ratio is now 0.5510).

The optimum hBN surface coverage by the $XeF_2$ and $KrF_2$ electron acceptors was determined by analysing their effect on the interface and their mutual interactions. Three different coverage levels were considered: ~1/3, ~2/3 and full molecular coverage with respect to the number of boron atoms. The ~1/3 coverage was selected because it avoided the unphysical covalent bonding of noble gas fluorides both with the hBN separator and between the neighbouring acceptor molecules.

## 4. Screening of TMD | hBN | OX Systems

In this work, we have focused on $h^+$ doping to TMDs. Suitable combinations of a TMD and an electron acceptor were identified based on the relative positions of their electronic states. The valence band maximum (VBM) and conduction band minimum (CBM), which were calculated from DFT calculations (see Table 2), provide an energetic criterion for estimating the expected direction of charge transfer and classifying the corresponding band alignment.

**Table 2.** The valence band maximum (VBM), conduction band minimum (CBM) and band gaps in units of eV for a range of TMDs and accessible good electron acceptors. hBN was added for comparison, its large band gap rendering it as a good dielectric separator in most van der Waals systems. The energy scales were aligned with respect to the He 1s peak.

| | hBN | $MoSe_2$ | $MoS_2$ | $WSe_2$ | $WS_2$ | $XeF_2$ | $KrF_2$ | $OsO_4$ |
|---|---|---|---|---|---|---|---|---|
| VBM | 9.86 | 10.37 | 9.76 | 10.8 | 9.98 | 6.55 | 6.58 | 6.8 |
| CBM | 14.14 | 11.83 | 11.43 | 12.09 | 11.82 | 9.8 | 9.45 | 10.81 |
| Band gap | 4.28 | 1.46 | 1.67 | 1.29 | 1.84 | 3.25 | 2.87 | 4.01 |

The expected direction of charge transfer depends strongly on the relative positions of the electronic states of the materials forming the heterostructure. Their band alignment is conventionally classified into three types[1]. In a type I band alignment, both the VBM and the CBM of one material lie within the band gap of the other material. Consequently, both electrons and holes preferentially localize in the same constituent. In a type II band alignment, the highest VBM and the lowest CBM belong to different materials but the highest VBM is below the lowest CBM, promoting the spatial separation of electrons and holes. Type III corresponds to a broken-gap alignment, in which the CBM of one material lies below the VBM of the other. Such an arrangement provides an energetic pathway for electrons to be transferred from the valence states of the electron donor to the initially unoccupied states of the acceptor. An example of the type III band alignment between $WSe_2$ and $KrF_2$ is shown in Figure 1.

The screening was therefore focused on identifying type III combinations with the largest energetic overlap between the donor and acceptor states. The calculated VBM and CBM positions were subsequently used to classify the investigated TMD/acceptor pairs according to their band-alignment type. The results are summarized in Table 3. Among the investigated combinations, the largest driving force for electron transfer (i.e. ultimately the difference of chemical potentials of OX and RED constituents[16]) was obtained for the $WSe_2/KrF_2$ pair. On the other hand, the smallest (but non-zero) one was computed for the $WS_2/XeF_2$ pair. To examine the influence of both the TMD and the electron acceptor, and to assess the predictive power of the proposed screening criterion for tungsten dichalcogenides, the $WS_2$ | hBN | $XeF_2$ and $WSe_2$ | hBN | $KrF_2$ heterostructures were selected for further analysis and calculation of the achievable CT.

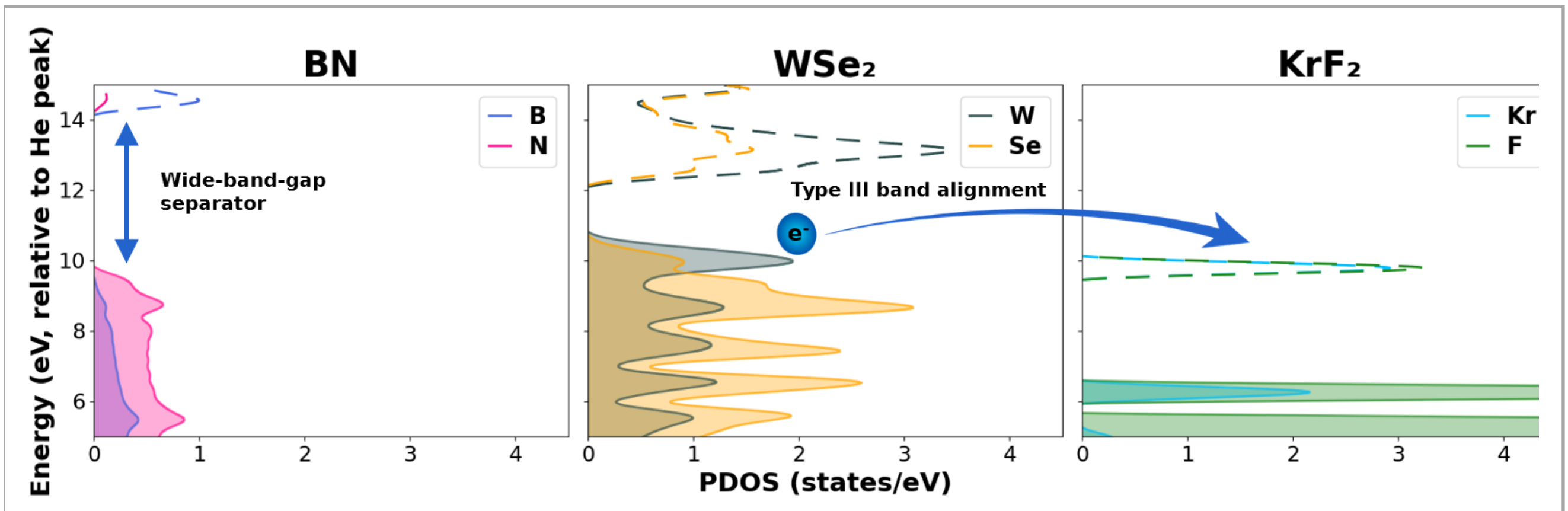

**Figure 1.** An example of projected density of states (PDOS) for a TMD | hBN | OX system. The colour-filled bands are the valence bands occupied by electrons, whereas dashed lines mark contours of the conduction bands. As can be seen from the positions of electronic bands for $WSe_2$ and $KrF_2$, CBM of $KrF_2$ lies below VBM of $WSe_2$, which indicates a type III band alignment and the electrons transfer from $WSe_2$ to $KrF_2$ (strong electron acceptor). The hBN PDOS shows its wide band gap encompassing key electronic features of OX and RED, thus providing its "neutral observer" and dielectric separator status.

**Table 3.** Comparison of VBM and CBM differences ($E_{OX}^{CBM}$-$E_{TMD}^{VBM}$) for TMD/OX systems in units of eV. Negative values represent systems with type III band alignment, where TMD valence band maximum lies above conduction bands minimum of the electron acceptor. The positive values result in null CT between RED and OX.

| | $XeF_2$ | $KrF_2$ | $OsO_4$ |
|---|---|---|---|
| $MoSe_2$ | -0.57 | -0.91 | -1.01 |
| $MoS_2$ | 0.04 | -0.31 | -0.95 |
| $WSe_2$ | -1.01 | -1.35 | -1.24 |
| $WS_2$ | -0.18 | -0.52 | -0.99 |

## 5. CT in Tungsten Dichalcogenide Heterostructures

The optimized structures of the $WS_2$ | hBN | $XeF_2$ and $WSe_2$ | hBN | $KrF_2$ heterostructures are shown in Figure 2. The net charge transfer associated with each constituent layer was estimated from its PDOS. Since integration of atom- and orbital-projected DOS does not generally reproduce the exact number of valence electrons because of the incomplete projection of the electronic density onto localized atomic orbitals, each PDOS was normalized to the expected population of valence electrons of the corresponding isolated layer. For layer $i$, the normalization factor was defined as

$$s_i = \frac{N_i}{\int_{-\infty}^{E_F} D_i(E)dE}$$

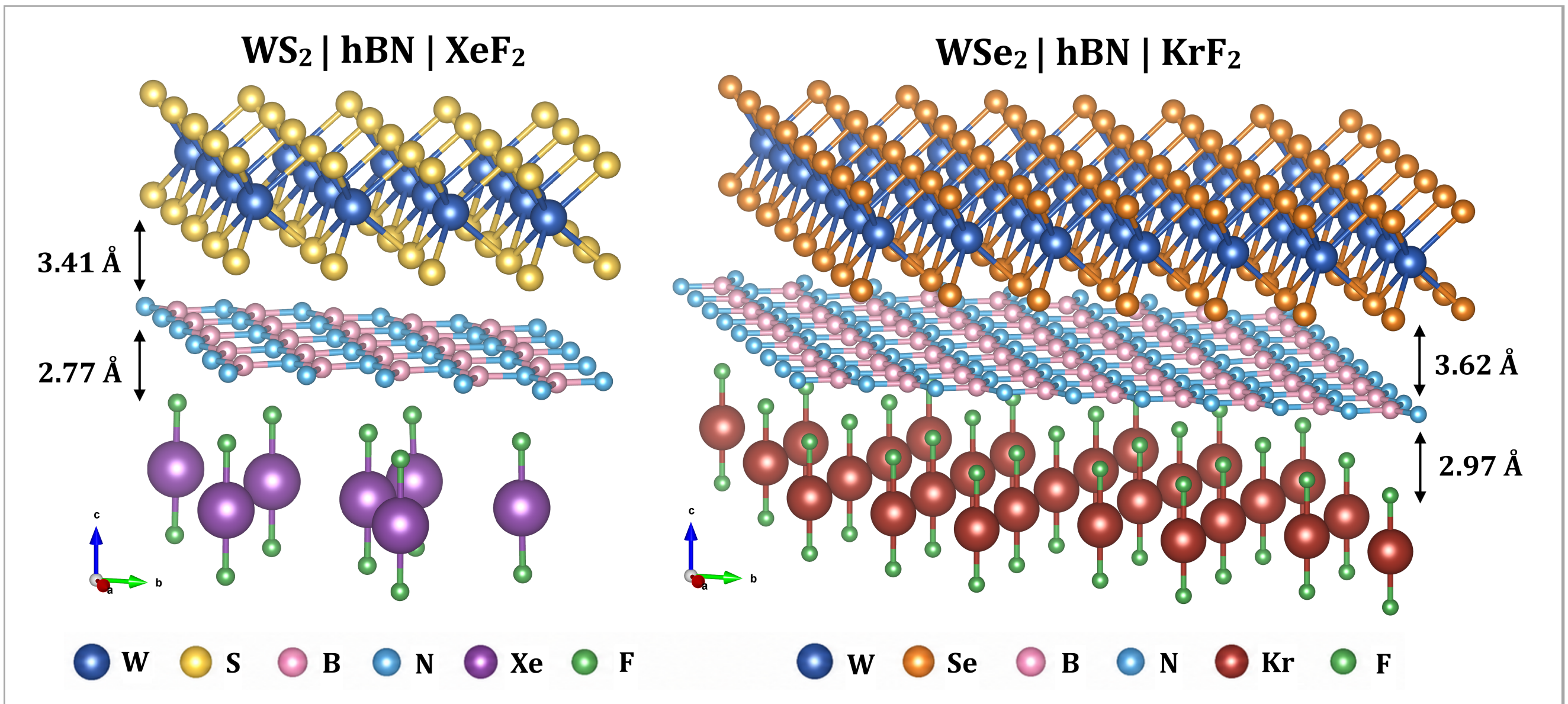

**Figure 2.** Side views of the $WS_2$| hBN | $XeF_2$ and $WSe_2$ | hBN | $KrF_2$ van der Waals heterostructures. The indicated interlayer distances are measured between the nearest chalcogen plane and the hBN layer and between the hBN layer and the nearest fluorine plane.

where $N_i$ is the number of valence electrons explicitly included in the employed pseudopotentials and $D_i$ is the density of states. The same normalization factor was subsequently applied to the PDOS of the corresponding constituent in the complete heterostructure:

$$\widetilde{D}_i(E) = s_i D_i(E)$$

And the change in the electronic occupation of layer *i* was calculated as

$$\Delta N_i = \int_{-\infty}^{E_F} \widetilde{D}_i^{HT}(E)dE - \int_{-\infty}^{E_F} \widetilde{D}_i(E)dE$$

This normalization compensates for the systematic loss inherent to orbital projections while preserving the relative changes in electronic occupation upon heterostructure formation.

The absolute net CT associated with the hBN separator remained below 0.004 $h^+$ per B atom in both heterostructures, indicating that hBN remains nearly charge neutral and effectively acts as a dielectric spacer between the TMD and the molecular acceptor layer while preserving a direct chemical reaction between them from occuring. In contrast, the calculated hole concentrations in the $WS_2$ and $WSe_2$ monolayers reached 0.23 $h^+$ and 0.35 $h^+$ per W atom, respectively. Comparison with the previous results summarized in Table 1 demonstrates that the proposed CC concept is theoretically capable of achieving large carrier concentrations comparable to electrostatic gating while exceeding those reported for other doping methods, without the application of an external electric field or any permanent chemical modification of the active TMD layer.

The larger charge transfer obtained for the $WSe_2$ | hBN | $KrF_2$ heterostructure in comparison to the $WS_2$ | hBN | $XeF_2$ is consistent with its larger initial energetic overlap between the $WSe_2$ valence band maximum and the unoccupied states of $KrF_2$, as well as the slightly larger OX:RED ratio for the former case. Interestingly, this is despite the fact that the equilibrium OX-RED separation in the latter is by ca. 0.5 Å larger for $WSe_2$-$KrF_2$ than for $WS_2$-$XeF_2$ layers. Consequently, the transferred charge is not proportional to the initial band offset determined for the isolated constituents. Instead, the final equilibrium charge distribution results from the interplay of several coupled structural (cf. SI), electronic and electrostatic effects. The transfer is primarily driven by the self-consistent equilibration of the electrochemical potentials of the constituent layers and progressively opposed by the electrostatic potential generated by the separated charges. The final magnitude of charge transfer is further influenced by dielectric screening within the heterostructure, the finite quantum capacitance of the constituent layers and polarization of the molecular acceptor, as well as the electronic coupling across the hBN separator[76].

## 6. Conclusions and Outlook

This work demonstrates the capabilities of the chemical capacitor concept[3][16] by applying its fundamental principles to experimentally accessible van der Waals heterostructures based on TMD monolayers. Despite the limitations of the present study related to the large unit cell size of the investigated heterostructures and the substantial computational cost associated with their structural relaxation, a theoretical proof of concept was successfully delivered. The achievable CT in a realistic, experimentally manufacturable system composed of TMD monolayers, dielectric hBN separator, and selected volatile fluoride oxidizers, reaches as much as 0.35 $h^+$ per TM atom in the $WSe_2$ | hBN | $KrF_2$ system. This value is comparable to that of 0.44 $h^+$/TM achieved via electric gating but the effect in our case comes from the difference of the chemical rather than electric potentials. We expect that the proposed heterostructures will serve as a foundation for the future experimental verification of the extent of the CT calculated here. Our preliminary results indicate that this approach could be extended also to the TMDs/RED (i.e. donor) systems such as RED=Li (see SI).

The proposed architecture is sufficiently promising to motivate further investigations of correlated electronic phenomena that may emerge at the achieved carrier concentrations, such as superconductivity. Beyond demonstrating exceptionally large interfacial CT, the CC systems embedded in van der Waals heterostructures provide a permanent and contactless route to engineering of carrier density in two-dimensional semiconductors and lay the foundation for future development of van der Waals devices.

## Acknowledgements

The authors are grateful to the Polish National Science Center for project OPUS Chem-Cap (2021/41/B/ST5/00195). Computations were performed at the Interdisciplinary Center for Mathematical and Computational Modeling of Warsaw University (grant SAPPHIRE GA83-34). This work is dedicated to the memory of Professor Edward Darżynkiewicz (1948–2024).

## References

[1] Briggs N *et al* 2019 A roadmap for electronic grade 2D materials *2D Mater.* **6** 022001.
[2] Wu R *et al* Synthesis, Modulation, and Application of Two-Dimensional TMD Heterostructures 2024 *Chemical Reviews 124* (17), 10112-10191.
[3] Manzeli S *et al* 2017 2D transition metal dichalcogenides. *Nat Rev Mater* 2, 17033.
[4] Wolański Ł, Ciszewski D, Szkudlarek P, Lorenzana J and Grochala W 2026 A chemical capacitor – its concept, functionalities and limits. *Phys. Chem. Chem. Phys.* 28 (4): 2767–2778.
[5] Szudlarek P G, Renskers C, Margine E R and Grochala W 2025 Superconducting Lithium Hydride in a Chemical Capacitor Setup: A Theoretical Study. *ChemPhysChem* 26: e202500013.
[6] Geim A K and Grigorieva I V 2013 Van der Waals heterostructures *Nature* Jul 25;499(7459):419-25.
[7] Jiamin X *et al* 2011 Scanning tunnelling microscopy and spectroscopy of ultra-flat graphene on hexagonal boron nitride *Nature Materials*, 10 (4), 282-285.
[8] Komsa H-P and Krasheninnikov A V. 2013 Electronic structures and optical properties of realistic transition metal dichalcogenide heterostructures from first principles *Phys. Rev. B* 88, 085318.
[9] Zhou H *et al* 2020 Controlling Exciton and Valley Dynamics in Two-Dimensional Heterostructures with Atomically Precise Interlayer Proximity *ACS Nano*, 14, 4618−4625.
[10] Mihyang Yu *et al* 2022 Electrical charge control of h-BN single photon sources *2D Mater.* 9 035020.
[11] Choi Y *et al* 2021 Multiterminal Transport Measurements of Multilayer InSe Encapsulated by hBN *ACS Applied Electronic Materials 3* (1), 163-169.
[12] Ki Kang K, Hyun Seok L and Young Hee L 2018 Synthesis of hexagonal boron nitride heterostructures for 2D van der Waals electronics. *Chem. Soc. Rev.* 47 (16): 6342–6369.
[13] Tanweer A *et al* 2020 Interplay of charge transfer and disorder in optoelectronic response in Graphene/hBN/MoS2 van der Waals heterostructures *2D Mater*. 7 025043.
[14] Yoon Y *et al* 2022 Charge Transfer Dynamics in MoSe2/hBN/WSe2 Heterostructures *Nano Letters 22* (24), 10140-10146.
[15] Fang H *et al* 2014 Strong interlayer coupling in van der Waals heterostructures built from single-layer chalcogenides. *Proc Natl Acad Sci U S A.* Apr 29;111(17):6198-202.
[16] Grzelak A, Lorenzana J, Grochala W 2021 Separation-Controlled Redox Reactions *Angew. Chem. Int. Ed.* 2021, *60*, 13892.
[17] Zhang X *et al* 2015 Phonon and Raman scattering of two-dimensional transition metal dichalcogenides from monolayer, multilayer to bulk material *Chem. Soc. Rev.* (2015) 44 (9): 2757–2785.
[18] Voiry D *et al* 2015 Phase engineering of transition metal dichalcogenides *Chem. Soc. Rev.* (2015) 44 (9): 2702–2712.
[19] Obaidulla S M *et al* Van der Waals 2D transition metal dichalcogenide/organic hybridized heterostructures: recent breakthroughs and emerging prospects of the device. *Nanoscale Horiz.* 2024; 9 (1): 44–92.
[20] Schaibley J, Yu H and Clark G *et al* 2016 Valleytronics in 2D materials. *Nat Rev Mater* 1, 16055.
[21] Shinde P A and Ariga K 2023 Two-Dimensional Nanoarchitectonics for Two-Dimensional Materials: Interfacial Engineering of Transition-Metal Dichalcogenides *Langmuir 39* (50), 18175-18186.
[22] Duan X *et al* 2015 Two-dimensional transition metal dichalcogenides as atomically thin semiconductors: opportunities and challenges. *Chem. Soc. Rev.* 2015; 44 (24): 8859–8876.

[23] Ruitao L *et al* 2015 Transition Metal Dichalcogenides and Beyond: Synthesis, Properties, and Applications of Single- and Few-Layer Nanosheets *Accounts of Chemical Research 48* (1), 56-64.
[24] Mak K F and Shan J 2016 Photonics and optoelectronics of 2D semiconductor transition metal dichalcogenides *Nature Photonics* 10, 216–226.
[25] Tan Ch and Zhang H 2015 Two-dimensional transition metal dichalcogenide nanosheet-based composites. *Chem. Soc. Rev.* 44 (9): 2713–2731.
[26] Zhang R *et al* 2016 Superconductivity in potassium-doped metallic polymorphs of MoS2 Nano Lett. 16 629–36.
[27] Hee Han G *et al* 2018 van der Waals Metallic Transition Metal Dichalcogenides *Chem Rev.* Jul 11;118(13):6297-6336.
[28] Li H-M *et al* 2015 Ultimate thin vertical p–n junction composed of two-dimensional layered molybdenum disulfide Nat. Commun. 6 6564.
[29] Zhang R, Drysdale D, Koutsos V and Cheung R 2017 Controlled Layer Thinning and p-Type Doping of $WSe_2$ by Vapor $XeF_2$ *Adv. Funct. Mater.* 2017, 27, 1702455.
[30] Yuan H *et al* 2013 Zeeman-type spin splitting controlled by an electric field Nat. Phys. 9 563–9.
[31] Costanzo D, Jo S, Berger H and Morpurgo A F 2016 Gateinduced superconductivity in atomically thin MoS2 crystals Nat. Nanotechnol. 11 339–44.
[32] Zhao B *et al* 2025 Gate-driven band modulation hyperdoping for high-performance p-type 2D semiconductor transistors *Science* 388, 1183-1188.
[33] Tongay S *et al* 2013 Broad-range modulation of light emission in two-dimensional semiconductors by molecular physisorption gating Nano Lett. 13 2831–6.
[34] Reed-Lingenfelter S N *et al* 2022 Compact Super Electron-Donor to Monolayer MoS2 *Nano Letters* 2022 22 (11), 4501-4508.
[35] Huang Y L *et al* 2018 The organic–2D transition metal dichalcogenide heterointerface. *Chem. Soc. Rev.* 47 (9): 3241–3264.
[36] Jing Y, Tan X, Zhou Z and Shen P W 2014 Tuning electronic and optical properties of MoS2 monolayer via molecular charge transfer *J. Mater. Chem. A* 2, 16892–16897.
[37] Zeng J *et al* 2021 Effects of Strain and Electric Field on Molecular Doping in MoSSe Ouyang *ACS Omega 6* (22), 14639-14647.
[38] Cai Y, Zhou H, Zhang G and Zhang Y-W 2016 Modulating Carrier Density and Transport Properties of MoS2 by Organic Molecular Doping and Defect Engineering *Chemistry of Materials* 2016 *28* (23), 8611-8621.
[39] Choudhury P, Ravavarapu L, Dekle R and Chowdhury S 2017 Modulating Electronic and Optical Properties of Monolayer MoS2 Using Nonbonded Phthalocyanine Molecules *The Journal of Physical Chemistry C* 2017 *121* (5), 2959-2967.
[40] Song Z. *et al* 2017 Electronic Properties of a 1D Intrinsic/p-Doped Heterojunction in a 2D Transition Metal Dichalcogenide Semiconductor *ACS Nano 11* (9), 9128-9135.
[41] Mouri S, Miyauchi Y and Matsuda K 2013 Tunable photoluminescence of monolayer MoS2 via chemical doping Nano Lett. 13 5944–8.
[42] Zhao P *et al* 2014 Air stable p-doping of $WSe_2$ by covalent functionalization ACS Nano 8 10808–14.
[43] Li X *et al* 2016 Isoelectronic tungsten doping in monolayer $MoSe_2$ for carrier type modulation Adv. Mater. 28 8240–7.
[44] Rhodes D *et al* 2017 Engineering the structural and electronic phases of $MoTe_2$ through W substitution Nano Lett. 17 1616–22.
[45] Gong Y *et al* 2014 Band gap engineering and layer-by-layer mapping of selenium-doped molybdenum disulfide Nano Lett. 14 442–9.
[46] Suh J *et al* 2014 Doping against the native propensity of $MoS_2$: degenerate hole doping by cation substitution Nano Lett. 14 6976–82.
[47] Yoshida S *et al* 2015 Microscopic basis for the band engineering of $Mo_{1-x}W_xS_2$-based heterojunction Sci. Rep. 5 14808.
[48] Li X *et al* 2017 Suppression of defects and deep levels using isoelectronic tungsten substitution in monolayer MoSe2 Adv. Funct. Mater. 27 1603850.
[49] Zhu CR *et al* 2018 TMD-based highly efficient electrocatalysts developed by combined computational and experimental approaches*. Chem Soc Rev.* Jun 18;47(12):4332-4356.
[50] Yan X *et al* 2021 Defect engineering and characterization of active sites for efficient electrocatalysis. *Nanoscale* 2021; 13 (6): 3327–3345.
[51] Xie L M 2015 Two-dimensional transition metal dichalcogenide alloys: preparation, characterization and applications. *Nanoscale* 2015; 7 (44): 18392–18401.
[52] Jin Y *et al* 2015 A Van Der Waals Homojunction: Ideal p–n Diode Behavior in $MoSe_2$ . *Adv. Mater*., 27: 5534-5540.
[53] Fang H *et al* 2013 Degenerate n-doping of few-layer transition metal dichalcogenides by potassium Nano Lett. 13 1991–5.
[54] Gao H *et al* 2020 Tuning Electrical Conductance of MoS2 Monolayers through Substitutional Doping *Nano Letters 20* (6), 4095-4101.
[55] Zhang K *et al* 2018 Tuning the Electronic and Photonic Properties of Monolayer $MoS_2$ via In Situ Rhenium Substitutional Doping *Adv. Funct. Mater.* 2018, 28, 1706950.
[56] Li H *et al* 2022 Reducing Contact Resistance and Boosting Device Performance of Monolayer $MoS_2$ by In Situ Fe Doping. *Adv. Mater.* 34, 2200885.
[57] Zhang Q *et al* 2020 Intercalation and exfoliation chemistries of transition metal dichalcogenides. *J. Mater. Chem. A* 2020; 8 (31): 15417–15444.
[58] Hromadova L, Martonak R and Tossatti E 2013 Structure change, layer sliding, and metallization in high-pressure $MoS_2$ *Phys. Rev. B* 87, 144105.
[59] Riflikova M, Martonak R and Tosatti E 2014 Pressure-induced gap closing and metallization of $MoSe_2$ and $MoTe_2$ *Phys. Rev. B* 90, 035108.
[60] Kohulak O, Martonak R and Tossati E 2015 High-pressure structure, decomposition, and superconductivity of *$MoS_2$ Phys. Rev. B* 91, 144113.
[61] Kohulak O and Martonak R 2022 New High-Pressure Phases of $MoSe_2$ and $MoTe_2$ *Phys. Rev. B* 95, 054105.
[62] Novoselov K S, Mishchenko A, Carvalho A and Castro Neto A H 2016 2D materials and van der Waals heterostructures. *Science* Jul 29;353(6298):aac9439.
[63] Liu Y *et al* 2016 Van der Waals heterostructures and devices. *Nat Rev Mater* 1, 16042.
[64] Hu Z *et al* 2022 Hernández-Martínez PL, *et al* Interfacial charge and energy transfer in van der Waals heterojunctions. *InfoMat*. 4(3):e12290.
[65] Sangwan V K and Hersam M C 2018 Electronic Transport in Two-Dimensional Materials. *Annual Review Physical Chemistry*. 69:299-325.

[66] Hong X *et al* 2014 Ultrafast charge transfer in atomically thin $MoS_2/WS_2$ heterostructures. *Nature Nanotech* 9, 682–686.
[67] Kresse G and Hafner J 1993 Ab initio molecular dynamics for liquid metals *Physical review B* 47, 558.
[68] Kresse G and Furthmüller J 1996 Efficiency of ab-initio total energy calculations for metals and semiconductors using a plane-wave basis set *Computational Materials Science* 6, 15.
[69] Kresse G and Furthmüller J 1996 Efficient iterative schemes for *ab initio* total-energy calculations using a plane-wave basis set *Physical review B* 54, 11169.
[70] Kresse G and Hafner J 1994 Norm-conserving and ultrasoft pseudopotentials for first-row and transition elements *Phys.: Condens. Matt.* 6, 8245.
[71] Blöchl P E 1994 Projector augmented-wave method *Physical review B* 50, 17953.
[72] Kresse G and Joubert D 1999 From ultrasoft pseudopotentials to the projector augmented-wave method *Physical review B* 59, 1758.
[73] Sun J, Ruzsinszky A and Perdew J P 2015 Strongly constrained and appropriately normed semilocal density functional *Phys. Rev. Lett.* 115 036402.
[74] Sabatini R, Gorni T and de Gironcoli S 2013 Nonlocal van der Waals density functional made simple and efficient Phys. Rev. B 87 041108(R).
[75] Bengtsson L 1999 Dipole correction for surface supercell calculations Phys. Rev. B 59 12301–12304.
[76] Jezierski D *et al* 2022 Charge doping to flat $AgF_2$ monolayers in a chemical capacitor setup. *Phys. Chem. Chem. Phys.* 2022; 24 (26): 15705–15717.
[77] Lazić P 2015 *CellMatch*: Combining two unit cells into a common supercell with minimal strain *Computer Physics Communications* 197, 324-334.

# Huge hole injection in tungsten dichalcogenide heterostructures without electric gating: a DFT study

**Dawid Ciszewski[1*] and Wojciech Grochala[1*]**

[1] Centre of New Technologies University of Warsaw, Żwirki i Wigury 93, 02-089 Warsaw, Poland

*Authors to whom any correspondence should be addressed.

## Supplementary Information

### S1. CIF data before and after structural relaxation for $WS_2$ | hBN | $XeF_2$ and $WSe_2$ | hBN | $KrF_2$ heterostructures

The starting structures for relaxation were constructed in such a way that the preoptimized $NgF_2$ molecules were placed on an opposed side of the preoptimized heterostructure with the BN layer and the TMD layer, thus forming the TMD | hBN | OX system. Therefore, the starting geometry reflects well the optimized TMD | hBN and $NgF_2$ geometries as well as the initial location of the $NgF_2$ molecules with approx. 1/3 coverage.

**$WS_2$ | hBN | $XeF_2$ before CT**

```
_cell_length_a                12.564592
_cell_length_b                12.564592
_cell_length_c                49.635239
_cell_angle_alpha             90.000000
_cell_angle_beta              90.000000
_cell_angle_gamma             60.000000
_cell_volume                  6786.057767
_space_group_name_H-M_alt     'P 1'
_space_group_IT_number        1

loop_
_space_group_symop_operation_xyz
  'x, y, z'

loop_
  _atom_site_label
  _atom_site_occupancy
  _atom_site_fract_x
  _atom_site_fract_y
  _atom_site_fract_z
  _atom_site_adp_type
  _atom_site_U_iso_or_equiv
  _atom_site_type_symbol
  W1     1.0   0.916620   0.916619   0.600724   Uiso  ? W
  W2     1.0   0.916610   0.666631   0.600919   Uiso  ? W
  W3     1.0   0.916610   0.416759   0.600919   Uiso  ? W
  W4     1.0   0.916620   0.166761   0.600725   Uiso  ? W
  W5     1.0   0.666631   0.916609   0.600919   Uiso  ? W
  W6     1.0   0.666667   0.666666   0.601002   Uiso  ? W
  W7     1.0   0.666631   0.416759   0.600919   Uiso  ? W
  W8     1.0   0.666606   0.166697   0.600843   Uiso  ? W
  W9     1.0   0.416759   0.916609   0.600919   Uiso  ? W
  W10    1.0   0.416759   0.666631   0.600919   Uiso  ? W
  W11    1.0   0.416674   0.416674   0.600956   Uiso  ? W
  W12    1.0   0.416675   0.166651   0.600956   Uiso  ? W
  W13    1.0   0.166761   0.916619   0.600724   Uiso  ? W
  W14    1.0   0.166697   0.666606   0.600843   Uiso  ? W
```

W15 1.0 0.166651 0.416674 0.600956 Uiso ? W
W16 1.0 0.166697 0.166697 0.600843 Uiso ? W
S1 1.0 0.833485 0.833485 0.632378 Uiso ? S
S2 1.0 0.833450 0.583275 0.632492 Uiso ? S
S3 1.0 0.833485 0.333030 0.632378 Uiso ? S
S4 1.0 0.833558 0.083221 0.632267 Uiso ? S
S5 1.0 0.583275 0.833449 0.632492 Uiso ? S
S6 1.0 0.583275 0.583275 0.632492 Uiso ? S
S7 1.0 0.583487 0.333174 0.632440 Uiso ? S
S8 1.0 0.583487 0.083338 0.632441 Uiso ? S
S9 1.0 0.333030 0.833485 0.632378 Uiso ? S
S10 1.0 0.333175 0.583487 0.632441 Uiso ? S
S11 1.0 0.333333 0.333333 0.632500 Uiso ? S
S12 1.0 0.333174 0.083338 0.632441 Uiso ? S
S13 1.0 0.083221 0.833558 0.632267 Uiso ? S
S14 1.0 0.083339 0.583487 0.632441 Uiso ? S
S15 1.0 0.083339 0.333174 0.632441 Uiso ? S
S16 1.0 0.083221 0.083221 0.632267 Uiso ? S
S17 1.0 0.833012 0.833012 0.569357 Uiso ? S
S18 1.0 0.833093 0.583453 0.569470 Uiso ? S
S19 1.0 0.833012 0.333976 0.569357 Uiso ? S
S20 1.0 0.832954 0.083523 0.569254 Uiso ? S
S21 1.0 0.583454 0.833093 0.569470 Uiso ? S
S22 1.0 0.583454 0.583453 0.569470 Uiso ? S
S23 1.0 0.583060 0.333632 0.569434 Uiso ? S
S24 1.0 0.583060 0.083308 0.569434 Uiso ? S
S25 1.0 0.333977 0.833012 0.569357 Uiso ? S
S26 1.0 0.333632 0.583060 0.569434 Uiso ? S
S27 1.0 0.333333 0.333333 0.569506 Uiso ? S
S28 1.0 0.333632 0.083308 0.569434 Uiso ? S
S29 1.0 0.083523 0.832954 0.569254 Uiso ? S
S30 1.0 0.083308 0.583060 0.569434 Uiso ? S
S31 1.0 0.083308 0.333632 0.569434 Uiso ? S
S32 1.0 0.083523 0.083523 0.569254 Uiso ? S
B1 1.0 0.933359 0.933358 0.502594 Uiso ? B
B2 1.0 0.933352 0.733455 0.504453 Uiso ? B
B3 1.0 0.933348 0.533326 0.505656 Uiso ? B
B4 1.0 0.933352 0.333193 0.504453 Uiso ? B
B5 1.0 0.933359 0.133283 0.502594 Uiso ? B
B6 1.0 0.733455 0.933352 0.504453 Uiso ? B
B7 1.0 0.733384 0.733385 0.505936 Uiso ? B
B8 1.0 0.733384 0.533231 0.505936 Uiso ? B
B9 1.0 0.733455 0.333193 0.504453 Uiso ? B
B10 1.0 0.733479 0.133261 0.503580 Uiso ? B
B11 1.0 0.533326 0.933348 0.505656 Uiso ? B
B12 1.0 0.533231 0.733385 0.505936 Uiso ? B
B13 1.0 0.533326 0.533326 0.505656 Uiso ? B
B14 1.0 0.533469 0.333263 0.505268 Uiso ? B
B15 1.0 0.533469 0.133268 0.505268 Uiso ? B
B16 1.0 0.333193 0.933352 0.504453 Uiso ? B
B17 1.0 0.333193 0.733455 0.504453 Uiso ? B
B18 1.0 0.333263 0.533470 0.505268 Uiso ? B
B19 1.0 0.333333 0.333334 0.505814 Uiso ? B
B20 1.0 0.333263 0.133268 0.505268 Uiso ? B
B21 1.0 0.133283 0.933358 0.502594 Uiso ? B
B22 1.0 0.133260 0.733479 0.503580 Uiso ? B
B23 1.0 0.133268 0.533470 0.505268 Uiso ? B
B24 1.0 0.133268 0.333263 0.505268 Uiso ? B
B25 1.0 0.133260 0.133261 0.503580 Uiso ? B
N1 1.0 0.866722 0.866722 0.503508 Uiso ? N
N2 1.0 0.866730 0.666747 0.505517 Uiso ? N
N3 1.0 0.866730 0.466523 0.505517 Uiso ? N
N4 1.0 0.866722 0.266556 0.503508 Uiso ? N
N5 1.0 0.866695 0.066653 0.502333 Uiso ? N

N6 1.0 0.666747 0.866731 0.505517 Uiso ? N
N7 1.0 0.666666 0.666667 0.506363 Uiso ? N
N8 1.0 0.666747 0.466523 0.505517 Uiso ? N
N9 1.0 0.666797 0.266560 0.504239 Uiso ? N
N10 1.0 0.666797 0.066643 0.504239 Uiso ? N
N11 1.0 0.466522 0.866731 0.505517 Uiso ? N
N12 1.0 0.466523 0.666747 0.505517 Uiso ? N
N13 1.0 0.466674 0.466675 0.505490 Uiso ? N
N14 1.0 0.466764 0.266618 0.505527 Uiso ? N
N15 1.0 0.466674 0.066652 0.505490 Uiso ? N
N16 1.0 0.266556 0.866722 0.503508 Uiso ? N
N17 1.0 0.266560 0.666798 0.504239 Uiso ? N
N18 1.0 0.266618 0.466764 0.505528 Uiso ? N
N19 1.0 0.266618 0.266618 0.505527 Uiso ? N
N20 1.0 0.266560 0.066643 0.504239 Uiso ? N
N21 1.0 0.066653 0.866695 0.502333 Uiso ? N
N22 1.0 0.066642 0.666798 0.504239 Uiso ? N
N23 1.0 0.066651 0.466675 0.505490 Uiso ? N
N24 1.0 0.066642 0.266560 0.504239 Uiso ? N
N25 1.0 0.066653 0.066653 0.502333 Uiso ? N
F1 1.0 0.928031 0.535565 0.445196 Uiso ? F
F2 1.0 0.928031 0.535565 0.365106 Uiso ? F
F3 1.0 0.737578 0.130899 0.445196 Uiso ? F
F4 1.0 0.737578 0.130899 0.365106 Uiso ? F
F5 1.0 0.535700 0.928371 0.445196 Uiso ? F
F6 1.0 0.535700 0.928371 0.365106 Uiso ? F
F7 1.0 0.130507 0.739432 0.445196 Uiso ? F
F8 1.0 0.130507 0.739432 0.365106 Uiso ? F
F9 1.0 0.536190 0.535679 0.365106 Uiso ? F
F10 1.0 0.536190 0.535679 0.445196 Uiso ? F
F11 1.0 0.333140 0.333500 0.445196 Uiso ? F
F12 1.0 0.130943 0.130761 0.445196 Uiso ? F
F13 1.0 0.130943 0.130761 0.365106 Uiso ? F
F14 1.0 0.333140 0.333500 0.365106 Uiso ? F
Xe1 1.0 0.928031 0.535565 0.405151 Uiso ? Xe
Xe2 1.0 0.737578 0.130899 0.405151 Uiso ? Xe
Xe3 1.0 0.535700 0.928371 0.405151 Uiso ? Xe
Xe4 1.0 0.130507 0.739432 0.405151 Uiso ? Xe
Xe5 1.0 0.536190 0.535679 0.405151 Uiso ? Xe
Xe6 1.0 0.333140 0.333500 0.405151 Uiso ? Xe
Xe7 1.0 0.130943 0.130761 0.405151 Uiso ? Xe

**$WSe_2$ | hBN | $KrF_2$ before CT**

_cell_length_a 22.616266
_cell_length_b 22.616266
_cell_length_c 38.172848
_cell_angle_alpha 90.000000
_cell_angle_beta 90.000000
_cell_angle_gamma 60.000000
_cell_volume 16909.353692
_space_group_name_H-M_alt 'P 1'
_space_group_IT_number 1

loop_
_space_group_symop_operation_xyz
'x, y, z'

loop_
_atom_site_label
_atom_site_occupancy

_atom_site_fract_x
_atom_site_fract_y
_atom_site_fract_z
_atom_site_adp_type
_atom_site_U_iso_or_equiv
_atom_site_type_symbol
W1 1.0 0.952381 0.952381 0.641842 Uiso ? W
W2 1.0 0.809524 0.952381 0.641714 Uiso ? W
W3 1.0 0.952381 0.809524 0.641799 Uiso ? W
W4 1.0 0.666667 0.952381 0.641738 Uiso ? W
W5 1.0 0.809524 0.809524 0.641736 Uiso ? W
W6 1.0 0.952381 0.666667 0.641843 Uiso ? W
W7 1.0 0.523809 0.952381 0.641846 Uiso ? W
W8 1.0 0.666667 0.809524 0.641723 Uiso ? W
W9 1.0 0.809524 0.666667 0.641807 Uiso ? W
W10 1.0 0.952381 0.523810 0.641893 Uiso ? W
W11 1.0 0.380952 0.952381 0.641918 Uiso ? W
W12 1.0 0.523809 0.809524 0.641728 Uiso ? W
W13 1.0 0.666667 0.666667 0.641731 Uiso ? W
W14 1.0 0.809524 0.523810 0.641834 Uiso ? W
W15 1.0 0.952381 0.380952 0.641871 Uiso ? W
W16 1.0 0.238095 0.952381 0.641965 Uiso ? W
W17 1.0 0.380952 0.809524 0.641815 Uiso ? W
W18 1.0 0.523809 0.666667 0.641737 Uiso ? W
W19 1.0 0.666667 0.523810 0.641822 Uiso ? W
W20 1.0 0.809524 0.380952 0.641862 Uiso ? W
W21 1.0 0.952381 0.238095 0.641860 Uiso ? W
W22 1.0 0.095238 0.952381 0.641963 Uiso ? W
W23 1.0 0.238095 0.809524 0.641929 Uiso ? W
W24 1.0 0.380952 0.666667 0.641833 Uiso ? W
W25 1.0 0.523809 0.523810 0.641878 Uiso ? W
W26 1.0 0.666667 0.380952 0.641939 Uiso ? W
W27 1.0 0.809524 0.238095 0.641882 Uiso ? W
W28 1.0 0.952381 0.095238 0.641879 Uiso ? W
W29 1.0 0.095238 0.809524 0.641912 Uiso ? W
W30 1.0 0.238095 0.666667 0.641875 Uiso ? W
W31 1.0 0.380952 0.523810 0.641887 Uiso ? W
W32 1.0 0.523809 0.380952 0.641971 Uiso ? W
W33 1.0 0.666667 0.238095 0.641921 Uiso ? W
W34 1.0 0.809524 0.095238 0.641807 Uiso ? W
W35 1.0 0.095238 0.666667 0.641851 Uiso ? W
W36 1.0 0.238095 0.523810 0.641864 Uiso ? W
W37 1.0 0.380952 0.380952 0.641956 Uiso ? W
W38 1.0 0.523809 0.238095 0.641970 Uiso ? W
W39 1.0 0.666667 0.095238 0.641804 Uiso ? W
W40 1.0 0.095238 0.523810 0.641889 Uiso ? W
W41 1.0 0.238095 0.380952 0.641967 Uiso ? W
W42 1.0 0.380952 0.238095 0.642052 Uiso ? W
W43 1.0 0.523809 0.095238 0.641936 Uiso ? W
W44 1.0 0.095238 0.380952 0.641947 Uiso ? W
W45 1.0 0.238095 0.238095 0.642058 Uiso ? W
W46 1.0 0.380952 0.095238 0.642053 Uiso ? W
W47 1.0 0.095238 0.238095 0.641943 Uiso ? W
W48 1.0 0.238095 0.095238 0.642027 Uiso ? W
W49 1.0 0.095238 0.095238 0.641945 Uiso ? W
Se1 1.0 0.904762 0.904762 0.598024 Uiso ? Se
Se2 1.0 0.761905 0.904762 0.597964 Uiso ? Se
Se3 1.0 0.904762 0.761905 0.598037 Uiso ? Se
Se4 1.0 0.619048 0.904762 0.598010 Uiso ? Se
Se5 1.0 0.761905 0.761905 0.598005 Uiso ? Se
Se6 1.0 0.904762 0.619048 0.598102 Uiso ? Se
Se7 1.0 0.476191 0.904762 0.598070 Uiso ? Se
Se8 1.0 0.619048 0.761905 0.597969 Uiso ? Se
Se9 1.0 0.761905 0.619048 0.598043 Uiso ? Se

Se10 1.0 0.904762 0.476190 0.598110 Uiso ? Se
Se11 1.0 0.333333 0.904762 0.598134 Uiso ? Se
Se12 1.0 0.476191 0.761905 0.597991 Uiso ? Se
Se13 1.0 0.619048 0.619048 0.598002 Uiso ? Se
Se14 1.0 0.761905 0.476190 0.598083 Uiso ? Se
Se15 1.0 0.904762 0.333333 0.598094 Uiso ? Se
Se16 1.0 0.190476 0.904762 0.598196 Uiso ? Se
Se17 1.0 0.333333 0.761905 0.598096 Uiso ? Se
Se18 1.0 0.476191 0.619048 0.598053 Uiso ? Se
Se19 1.0 0.619048 0.476190 0.598127 Uiso ? Se
Se20 1.0 0.761905 0.333333 0.598137 Uiso ? Se
Se21 1.0 0.904762 0.190476 0.598104 Uiso ? Se
Se22 1.0 0.047619 0.904762 0.598155 Uiso ? Se
Se23 1.0 0.190476 0.761905 0.598151 Uiso ? Se
Se24 1.0 0.333333 0.619048 0.598102 Uiso ? Se
Se25 1.0 0.476191 0.476190 0.598156 Uiso ? Se
Se26 1.0 0.619048 0.333333 0.598194 Uiso ? Se
Se27 1.0 0.761905 0.190476 0.598112 Uiso ? Se
Se28 1.0 0.904762 0.047619 0.598081 Uiso ? Se
Se29 1.0 0.047619 0.761905 0.598097 Uiso ? Se
Se30 1.0 0.190476 0.619048 0.598096 Uiso ? Se
Se31 1.0 0.333333 0.476190 0.598131 Uiso ? Se
Se32 1.0 0.476191 0.333333 0.598205 Uiso ? Se
Se33 1.0 0.619048 0.190476 0.598137 Uiso ? Se
Se34 1.0 0.761905 0.047619 0.598007 Uiso ? Se
Se35 1.0 0.047619 0.619048 0.598100 Uiso ? Se
Se36 1.0 0.190476 0.476190 0.598132 Uiso ? Se
Se37 1.0 0.333333 0.333333 0.598221 Uiso ? Se
Se38 1.0 0.476191 0.190476 0.598224 Uiso ? Se
Se39 1.0 0.619048 0.047619 0.598059 Uiso ? Se
Se40 1.0 0.047619 0.476190 0.598148 Uiso ? Se
Se41 1.0 0.190476 0.333333 0.598223 Uiso ? Se
Se42 1.0 0.333333 0.190476 0.598297 Uiso ? Se
Se43 1.0 0.476191 0.047619 0.598190 Uiso ? Se
Se44 1.0 0.047619 0.333333 0.598151 Uiso ? Se
Se45 1.0 0.190476 0.190476 0.598244 Uiso ? Se
Se46 1.0 0.333333 0.047619 0.598243 Uiso ? Se
Se47 1.0 0.047619 0.190476 0.598139 Uiso ? Se
Se48 1.0 0.190476 0.047619 0.598213 Uiso ? Se
Se49 1.0 0.047619 0.047619 0.598164 Uiso ? Se
Se50 1.0 0.904762 0.904762 0.685545 Uiso ? Se
Se51 1.0 0.761905 0.904762 0.685485 Uiso ? Se
Se52 1.0 0.904762 0.761905 0.685558 Uiso ? Se
Se53 1.0 0.619048 0.904762 0.685531 Uiso ? Se
Se54 1.0 0.761905 0.761905 0.685526 Uiso ? Se
Se55 1.0 0.904762 0.619048 0.685623 Uiso ? Se
Se56 1.0 0.476191 0.904762 0.685593 Uiso ? Se
Se57 1.0 0.619048 0.761905 0.685491 Uiso ? Se
Se58 1.0 0.761905 0.619048 0.685564 Uiso ? Se
Se59 1.0 0.904762 0.476190 0.685632 Uiso ? Se
Se60 1.0 0.333333 0.904762 0.685655 Uiso ? Se
Se61 1.0 0.476191 0.761905 0.685512 Uiso ? Se
Se62 1.0 0.619048 0.619048 0.685523 Uiso ? Se
Se63 1.0 0.761905 0.476190 0.685604 Uiso ? Se
Se64 1.0 0.904762 0.333333 0.685615 Uiso ? Se
Se65 1.0 0.190476 0.904762 0.685717 Uiso ? Se
Se66 1.0 0.333333 0.761905 0.685617 Uiso ? Se
Se67 1.0 0.476191 0.619048 0.685574 Uiso ? Se
Se68 1.0 0.619048 0.476190 0.685648 Uiso ? Se
Se69 1.0 0.761905 0.333333 0.685658 Uiso ? Se
Se70 1.0 0.904762 0.190476 0.685625 Uiso ? Se
Se71 1.0 0.047619 0.904762 0.685677 Uiso ? Se
Se72 1.0 0.190476 0.761905 0.685674 Uiso ? Se
Se73 1.0 0.333333 0.619048 0.685624 Uiso ? Se

Se74 1.0 0.476191 0.476190 0.685677 Uiso ? Se
Se75 1.0 0.619048 0.333333 0.685717 Uiso ? Se
Se76 1.0 0.761905 0.190476 0.685634 Uiso ? Se
Se77 1.0 0.904762 0.047619 0.685604 Uiso ? Se
Se78 1.0 0.047619 0.761905 0.685620 Uiso ? Se
Se79 1.0 0.190476 0.619048 0.685617 Uiso ? Se
Se80 1.0 0.333333 0.476190 0.685652 Uiso ? Se
Se81 1.0 0.476191 0.333333 0.685726 Uiso ? Se
Se82 1.0 0.619048 0.190476 0.685658 Uiso ? Se
Se83 1.0 0.761905 0.047619 0.685528 Uiso ? Se
Se84 1.0 0.047619 0.619048 0.685621 Uiso ? Se
Se85 1.0 0.190476 0.476190 0.685653 Uiso ? Se
Se86 1.0 0.333333 0.333333 0.685744 Uiso ? Se
Se87 1.0 0.476191 0.190476 0.685745 Uiso ? Se
Se88 1.0 0.619048 0.047619 0.685580 Uiso ? Se
Se89 1.0 0.047619 0.476190 0.685670 Uiso ? Se
Se90 1.0 0.190476 0.333333 0.685744 Uiso ? Se
Se91 1.0 0.333333 0.190476 0.685818 Uiso ? Se
Se92 1.0 0.476191 0.047619 0.685712 Uiso ? Se
Se93 1.0 0.047619 0.333333 0.685672 Uiso ? Se
Se94 1.0 0.190476 0.190476 0.685766 Uiso ? Se
Se95 1.0 0.333333 0.047619 0.685764 Uiso ? Se
Se96 1.0 0.047619 0.190476 0.685661 Uiso ? Se
Se97 1.0 0.190476 0.047619 0.685734 Uiso ? Se
Se98 1.0 0.047619 0.047619 0.685686 Uiso ? Se
B1 1.0 0.925926 0.925926 0.479237 Uiso ? B
B2 1.0 0.814815 0.925926 0.479137 Uiso ? B
B3 1.0 0.925926 0.814815 0.479214 Uiso ? B
B4 1.0 0.703704 0.925926 0.479141 Uiso ? B
B5 1.0 0.814815 0.814815 0.479163 Uiso ? B
B6 1.0 0.925926 0.703704 0.479260 Uiso ? B
B7 1.0 0.592593 0.925926 0.479214 Uiso ? B
B8 1.0 0.703704 0.814815 0.479152 Uiso ? B
B9 1.0 0.814815 0.703704 0.479230 Uiso ? B
B10 1.0 0.925926 0.592593 0.479320 Uiso ? B
B11 1.0 0.481482 0.925926 0.479286 Uiso ? B
B12 1.0 0.592593 0.814815 0.479149 Uiso ? B
B13 1.0 0.703704 0.703704 0.479170 Uiso ? B
B14 1.0 0.814815 0.592593 0.479268 Uiso ? B
B15 1.0 0.925926 0.481481 0.479328 Uiso ? B
B16 1.0 0.370370 0.925926 0.479342 Uiso ? B
B17 1.0 0.481482 0.814815 0.479177 Uiso ? B
B18 1.0 0.592593 0.703704 0.479130 Uiso ? B
B19 1.0 0.703704 0.592593 0.479207 Uiso ? B
B20 1.0 0.814815 0.481481 0.479281 Uiso ? B
B21 1.0 0.925926 0.370370 0.479301 Uiso ? B
B22 1.0 0.259259 0.925926 0.479402 Uiso ? B
B23 1.0 0.370370 0.814815 0.479265 Uiso ? B
B24 1.0 0.481482 0.703704 0.479171 Uiso ? B
B25 1.0 0.592593 0.592593 0.479208 Uiso ? B
B26 1.0 0.703704 0.481481 0.479289 Uiso ? B
B27 1.0 0.814815 0.370370 0.479307 Uiso ? B
B28 1.0 0.925926 0.259259 0.479297 Uiso ? B
B29 1.0 0.148148 0.925926 0.479431 Uiso ? B
B30 1.0 0.259259 0.814815 0.479368 Uiso ? B
B31 1.0 0.370370 0.703704 0.479269 Uiso ? B
B32 1.0 0.481482 0.592593 0.479265 Uiso ? B
B33 1.0 0.592593 0.481481 0.479342 Uiso ? B
B34 1.0 0.703704 0.370370 0.479373 Uiso ? B
B35 1.0 0.814815 0.259259 0.479328 Uiso ? B
B36 1.0 0.925926 0.148148 0.479312 Uiso ? B
B37 1.0 0.037037 0.925926 0.479370 Uiso ? B
B38 1.0 0.148148 0.814815 0.479392 Uiso ? B
B39 1.0 0.259259 0.703704 0.479333 Uiso ? B

B40 1.0 0.370370 0.592593 0.479306 Uiso ? B
B41 1.0 0.481482 0.481481 0.479370 Uiso ? B
B42 1.0 0.592593 0.370370 0.479424 Uiso ? B
B43 1.0 0.703704 0.259259 0.479374 Uiso ? B
B44 1.0 0.814815 0.148148 0.479288 Uiso ? B
B45 1.0 0.925926 0.037037 0.479292 Uiso ? B
B46 1.0 0.037037 0.814815 0.479315 Uiso ? B
B47 1.0 0.148148 0.703704 0.479323 Uiso ? B
B48 1.0 0.259259 0.592593 0.479301 Uiso ? B
B49 1.0 0.370370 0.481481 0.479348 Uiso ? B
B50 1.0 0.481482 0.370370 0.479426 Uiso ? B
B51 1.0 0.592593 0.259259 0.479405 Uiso ? B
B52 1.0 0.703704 0.148148 0.479280 Uiso ? B
B53 1.0 0.814815 0.037037 0.479196 Uiso ? B
B54 1.0 0.037037 0.703704 0.479283 Uiso ? B
B55 1.0 0.148148 0.592593 0.479294 Uiso ? B
B56 1.0 0.259259 0.481481 0.479330 Uiso ? B
B57 1.0 0.370370 0.370370 0.479418 Uiso ? B
B58 1.0 0.481482 0.259259 0.479448 Uiso ? B
B59 1.0 0.592593 0.148148 0.479335 Uiso ? B
B60 1.0 0.703704 0.037037 0.479178 Uiso ? B
B61 1.0 0.037037 0.592593 0.479314 Uiso ? B
B62 1.0 0.148148 0.481481 0.479348 Uiso ? B
B63 1.0 0.259259 0.370370 0.479431 Uiso ? B
B64 1.0 0.370370 0.259259 0.479508 Uiso ? B
B65 1.0 0.481482 0.148148 0.479452 Uiso ? B
B66 1.0 0.592593 0.037037 0.479273 Uiso ? B
B67 1.0 0.037037 0.481481 0.479361 Uiso ? B
B68 1.0 0.148148 0.370370 0.479424 Uiso ? B
B69 1.0 0.259259 0.259259 0.479524 Uiso ? B
B70 1.0 0.370370 0.148148 0.479543 Uiso ? B
B71 1.0 0.481482 0.037037 0.479403 Uiso ? B
B72 1.0 0.037037 0.370370 0.479365 Uiso ? B
B73 1.0 0.148148 0.259259 0.479451 Uiso ? B
B74 1.0 0.259259 0.148148 0.479528 Uiso ? B
B75 1.0 0.370370 0.037037 0.479469 Uiso ? B
B76 1.0 0.037037 0.259259 0.479344 Uiso ? B
B77 1.0 0.148148 0.148148 0.479436 Uiso ? B
B78 1.0 0.259259 0.037037 0.479463 Uiso ? B
B79 1.0 0.037037 0.148148 0.479352 Uiso ? B
B80 1.0 0.148148 0.037037 0.479430 Uiso ? B
B81 1.0 0.037037 0.037037 0.479381 Uiso ? B
N1 1.0 0.962963 0.962963 0.479242 Uiso ? N
N2 1.0 0.851852 0.962963 0.479121 Uiso ? N
N3 1.0 0.962963 0.851852 0.479195 Uiso ? N
N4 1.0 0.740741 0.962963 0.479081 Uiso ? N
N5 1.0 0.851852 0.851852 0.479107 Uiso ? N
N6 1.0 0.962963 0.740741 0.479194 Uiso ? N
N7 1.0 0.629630 0.962963 0.479145 Uiso ? N
N8 1.0 0.740741 0.851852 0.479089 Uiso ? N
N9 1.0 0.851852 0.740741 0.479163 Uiso ? N
N10 1.0 0.962963 0.629630 0.479245 Uiso ? N
N11 1.0 0.518518 0.962963 0.479243 Uiso ? N
N12 1.0 0.629630 0.851852 0.479108 Uiso ? N
N13 1.0 0.740741 0.740741 0.479126 Uiso ? N
N14 1.0 0.851852 0.629630 0.479222 Uiso ? N
N15 1.0 0.962963 0.518519 0.479282 Uiso ? N
N16 1.0 0.407407 0.962963 0.479307 Uiso ? N
N17 1.0 0.518518 0.851852 0.479136 Uiso ? N
N18 1.0 0.629630 0.740741 0.479082 Uiso ? N
N19 1.0 0.740741 0.629630 0.479156 Uiso ? N
N20 1.0 0.851852 0.518519 0.479236 Uiso ? N
N21 1.0 0.962963 0.407407 0.479268 Uiso ? N
N22 1.0 0.296296 0.962963 0.479344 Uiso ? N

N23 1.0 0.407407 0.851852 0.479194 Uiso ? N
N24 1.0 0.518518 0.740741 0.479086 Uiso ? N
N25 1.0 0.629630 0.629630 0.479114 Uiso ? N
N26 1.0 0.740741 0.518519 0.479199 Uiso ? N
N27 1.0 0.851852 0.407407 0.479233 Uiso ? N
N28 1.0 0.962963 0.296296 0.479244 Uiso ? N
N29 1.0 0.185185 0.962963 0.479368 Uiso ? N
N30 1.0 0.296296 0.851852 0.479287 Uiso ? N
N31 1.0 0.407407 0.740741 0.479168 Uiso ? N
N32 1.0 0.518518 0.629630 0.479148 Uiso ? N
N33 1.0 0.629630 0.518519 0.479222 Uiso ? N
N34 1.0 0.740741 0.407407 0.479268 Uiso ? N
N35 1.0 0.851852 0.296296 0.479248 Uiso ? N
N36 1.0 0.962963 0.185185 0.479255 Uiso ? N
N37 1.0 0.074074 0.962963 0.479344 Uiso ? N
N38 1.0 0.185185 0.851852 0.479348 Uiso ? N
N39 1.0 0.296296 0.740741 0.479266 Uiso ? N
N40 1.0 0.407407 0.629630 0.479217 Uiso ? N
N41 1.0 0.518518 0.518519 0.479270 Uiso ? N
N42 1.0 0.629630 0.407407 0.479331 Uiso ? N
N43 1.0 0.740741 0.296296 0.479305 Uiso ? N
N44 1.0 0.851852 0.185185 0.479248 Uiso ? N
N45 1.0 0.962963 0.074074 0.479270 Uiso ? N
N46 1.0 0.074074 0.851852 0.479308 Uiso ? N
N47 1.0 0.185185 0.740741 0.479293 Uiso ? N
N48 1.0 0.296296 0.629630 0.479249 Uiso ? N
N49 1.0 0.407407 0.518519 0.479279 Uiso ? N
N50 1.0 0.518518 0.407407 0.479353 Uiso ? N
N51 1.0 0.629630 0.296296 0.479349 Uiso ? N
N52 1.0 0.740741 0.185185 0.479254 Uiso ? N
N53 1.0 0.851852 0.074074 0.479194 Uiso ? N
N54 1.0 0.074074 0.740741 0.479248 Uiso ? N
N55 1.0 0.185185 0.629630 0.479239 Uiso ? N
N56 1.0 0.296296 0.518519 0.479256 Uiso ? N
N57 1.0 0.407407 0.407407 0.479334 Uiso ? N
N58 1.0 0.518518 0.296296 0.479370 Uiso ? N
N59 1.0 0.629630 0.185185 0.479279 Uiso ? N
N60 1.0 0.740741 0.074074 0.479148 Uiso ? N
N61 1.0 0.074074 0.629630 0.479236 Uiso ? N
N62 1.0 0.185185 0.518519 0.479255 Uiso ? N
N63 1.0 0.296296 0.407407 0.479326 Uiso ? N
N64 1.0 0.407407 0.296296 0.479400 Uiso ? N
N65 1.0 0.518518 0.185185 0.479355 Uiso ? N
N66 1.0 0.629630 0.074074 0.479196 Uiso ? N
N67 1.0 0.074074 0.518519 0.479282 Uiso ? N
N68 1.0 0.185185 0.407407 0.479339 Uiso ? N
N69 1.0 0.296296 0.296296 0.479433 Uiso ? N
N70 1.0 0.407407 0.185185 0.479452 Uiso ? N
N71 1.0 0.518518 0.074074 0.479322 Uiso ? N
N72 1.0 0.074074 0.407407 0.479323 Uiso ? N
N73 1.0 0.185185 0.296296 0.479409 Uiso ? N
N74 1.0 0.296296 0.185185 0.479483 Uiso ? N
N75 1.0 0.407407 0.074074 0.479423 Uiso ? N
N76 1.0 0.074074 0.296296 0.479321 Uiso ? N
N77 1.0 0.185185 0.185185 0.479413 Uiso ? N
N78 1.0 0.296296 0.074074 0.479434 Uiso ? N
N79 1.0 0.074074 0.185185 0.479310 Uiso ? N
N80 1.0 0.185185 0.074074 0.479384 Uiso ? N
N81 1.0 0.074074 0.074074 0.479329 Uiso ? N
Kr1 1.0 0.925926 0.925926 0.364066 Uiso ? Kr
Kr2 1.0 0.814815 0.814815 0.364066 Uiso ? Kr
Kr3 1.0 0.592593 0.925926 0.364066 Uiso ? Kr
Kr4 1.0 0.925926 0.592593 0.364066 Uiso ? Kr
Kr5 1.0 0.703704 0.703704 0.364066 Uiso ? Kr

Kr6 1.0 0.481482 0.814815 0.364066 Uiso ? Kr
Kr7 1.0 0.814815 0.481481 0.364066 Uiso ? Kr
Kr8 1.0 0.259259 0.925926 0.364066 Uiso ? Kr
Kr9 1.0 0.592593 0.592593 0.364066 Uiso ? Kr
Kr10 1.0 0.925926 0.259259 0.364066 Uiso ? Kr
Kr11 1.0 0.370370 0.703704 0.364066 Uiso ? Kr
Kr12 1.0 0.703704 0.370370 0.364066 Uiso ? Kr
Kr13 1.0 0.148148 0.814815 0.364066 Uiso ? Kr
Kr14 1.0 0.481482 0.481481 0.364066 Uiso ? Kr
Kr15 1.0 0.814815 0.148148 0.364066 Uiso ? Kr
Kr16 1.0 0.259259 0.592593 0.364066 Uiso ? Kr
Kr17 1.0 0.592593 0.259259 0.364066 Uiso ? Kr
Kr18 1.0 0.037037 0.703704 0.364066 Uiso ? Kr
Kr19 1.0 0.370370 0.370370 0.364066 Uiso ? Kr
Kr20 1.0 0.703704 0.037037 0.364066 Uiso ? Kr
Kr21 1.0 0.148148 0.481481 0.364066 Uiso ? Kr
Kr22 1.0 0.481482 0.148148 0.364066 Uiso ? Kr
Kr23 1.0 0.259259 0.259259 0.364066 Uiso ? Kr
Kr24 1.0 0.037037 0.370370 0.364066 Uiso ? Kr
Kr25 1.0 0.370370 0.037037 0.364066 Uiso ? Kr
Kr26 1.0 0.148148 0.148148 0.364066 Uiso ? Kr
Kr27 1.0 0.037037 0.037037 0.364066 Uiso ? Kr
F1 1.0 0.925926 0.925926 0.413772 Uiso ? F
F2 1.0 0.814815 0.814815 0.413772 Uiso ? F
F3 1.0 0.592593 0.925926 0.413772 Uiso ? F
F4 1.0 0.925926 0.592593 0.413772 Uiso ? F
F5 1.0 0.703704 0.703704 0.413772 Uiso ? F
F6 1.0 0.481482 0.814815 0.413772 Uiso ? F
F7 1.0 0.814815 0.481481 0.413772 Uiso ? F
F8 1.0 0.259259 0.925926 0.413772 Uiso ? F
F9 1.0 0.592593 0.592593 0.413772 Uiso ? F
F10 1.0 0.925926 0.259259 0.413772 Uiso ? F
F11 1.0 0.370370 0.703704 0.413772 Uiso ? F
F12 1.0 0.703704 0.370370 0.413772 Uiso ? F
F13 1.0 0.148148 0.814815 0.413772 Uiso ? F
F14 1.0 0.481482 0.481481 0.413772 Uiso ? F
F15 1.0 0.814815 0.148148 0.413772 Uiso ? F
F16 1.0 0.259259 0.592593 0.413772 Uiso ? F
F17 1.0 0.592593 0.259259 0.413772 Uiso ? F
F18 1.0 0.037037 0.703704 0.413772 Uiso ? F
F19 1.0 0.370370 0.370370 0.413772 Uiso ? F
F20 1.0 0.703704 0.037037 0.413772 Uiso ? F
F21 1.0 0.148148 0.481481 0.413772 Uiso ? F
F22 1.0 0.481482 0.148148 0.413772 Uiso ? F
F23 1.0 0.259259 0.259259 0.413772 Uiso ? F
F24 1.0 0.037037 0.370370 0.413772 Uiso ? F
F25 1.0 0.370370 0.037037 0.413772 Uiso ? F
F26 1.0 0.148148 0.148148 0.413772 Uiso ? F
F27 1.0 0.037037 0.037037 0.413772 Uiso ? F
F28 1.0 0.925926 0.925926 0.314360 Uiso ? F
F29 1.0 0.814815 0.814815 0.314360 Uiso ? F
F30 1.0 0.592593 0.925926 0.314360 Uiso ? F
F31 1.0 0.925926 0.592593 0.314360 Uiso ? F
F32 1.0 0.703704 0.703704 0.314360 Uiso ? F
F33 1.0 0.481482 0.814815 0.314360 Uiso ? F
F34 1.0 0.814815 0.481481 0.314360 Uiso ? F
F35 1.0 0.259259 0.925926 0.314360 Uiso ? F
F36 1.0 0.592593 0.592593 0.314360 Uiso ? F
F37 1.0 0.925926 0.259259 0.314360 Uiso ? F
F38 1.0 0.370370 0.703704 0.314360 Uiso ? F
F39 1.0 0.703704 0.370370 0.314360 Uiso ? F
F40 1.0 0.148148 0.814815 0.314360 Uiso ? F
F41 1.0 0.481482 0.481481 0.314360 Uiso ? F
F42 1.0 0.814815 0.148148 0.314360 Uiso ? F

F43 1.0 0.259259 0.592593 0.314360 Uiso ? F
F44 1.0 0.592593 0.259259 0.314360 Uiso ? F
F45 1.0 0.037037 0.703704 0.314360 Uiso ? F
F46 1.0 0.370370 0.370370 0.314360 Uiso ? F
F47 1.0 0.703704 0.037037 0.314360 Uiso ? F
F48 1.0 0.148148 0.481481 0.314360 Uiso ? F
F49 1.0 0.481482 0.148148 0.314360 Uiso ? F
F50 1.0 0.259259 0.259259 0.314360 Uiso ? F
F51 1.0 0.037037 0.370370 0.314360 Uiso ? F
F52 1.0 0.370370 0.037037 0.314360 Uiso ? F
F53 1.0 0.148148 0.148148 0.314360 Uiso ? F
F54 1.0 0.037037 0.037037 0.314360 Uiso ? F

## $WS_2$ | hBN | $XeF_2$ after CT

_cell_length_a 12.564592
_cell_length_b 12.564592
_cell_length_c 49.635239
_cell_angle_alpha 90.000000
_cell_angle_beta 90.000000
_cell_angle_gamma 60.000000
_cell_volume 6786.057767
_space_group_name_H-M_alt 'P 1'
_space_group_IT_number 1

loop_
_space_group_symop_operation_xyz
'x, y, z'

loop_
_atom_site_label
_atom_site_occupancy
_atom_site_fract_x
_atom_site_fract_y
_atom_site_fract_z
_atom_site_adp_type
_atom_site_U_iso_or_equiv
_atom_site_type_symbol
W1 1.0 0.916714 0.916588 0.602180 Uiso ? W
W2 1.0 0.916712 0.666587 0.602290 Uiso ? W
W3 1.0 0.916713 0.416665 0.602290 Uiso ? W
W4 1.0 0.916714 0.166661 0.602179 Uiso ? W
W5 1.0 0.666712 0.916586 0.602289 Uiso ? W
W6 1.0 0.666741 0.666612 0.602338 Uiso ? W
W7 1.0 0.666714 0.416665 0.602289 Uiso ? W
W8 1.0 0.666686 0.166638 0.602243 Uiso ? W
W9 1.0 0.416794 0.916584 0.602291 Uiso ? W
W10 1.0 0.416795 0.666584 0.602291 Uiso ? W
W11 1.0 0.416741 0.416611 0.602307 Uiso ? W
W12 1.0 0.416740 0.166610 0.602307 Uiso ? W
W13 1.0 0.166789 0.916587 0.602181 Uiso ? W
W14 1.0 0.166767 0.666559 0.602247 Uiso ? W
W15 1.0 0.166738 0.416612 0.602309 Uiso ? W
W16 1.0 0.166766 0.166637 0.602246 Uiso ? W
S1 1.0 0.833491 0.833368 0.633759 Uiso ? S
S2 1.0 0.833472 0.583245 0.633829 Uiso ? S
S3 1.0 0.833492 0.333105 0.633758 Uiso ? S
S4 1.0 0.833515 0.083226 0.633692 Uiso ? S
S5 1.0 0.583377 0.833343 0.633829 Uiso ? S
S6 1.0 0.583378 0.583242 0.633829 Uiso ? S
S7 1.0 0.583485 0.333192 0.633795 Uiso ? S

S8 1.0 0.583483 0.083288 0.633795 Uiso ? S
S9 1.0 0.333235 0.833364 0.633760 Uiso ? S
S10 1.0 0.333326 0.583350 0.633797 Uiso ? S
S11 1.0 0.333408 0.333277 0.633831 Uiso ? S
S12 1.0 0.333324 0.083286 0.633797 Uiso ? S
S13 1.0 0.083348 0.833391 0.633694 Uiso ? S
S14 1.0 0.083413 0.583354 0.633797 Uiso ? S
S15 1.0 0.083413 0.333194 0.633797 Uiso ? S
S16 1.0 0.083347 0.083225 0.633693 Uiso ? S
S17 1.0 0.833226 0.833101 0.570727 Uiso ? S
S18 1.0 0.833282 0.583342 0.570792 Uiso ? S
S19 1.0 0.833227 0.333638 0.570726 Uiso ? S
S20 1.0 0.833171 0.083396 0.570668 Uiso ? S
S21 1.0 0.583465 0.833156 0.570792 Uiso ? S
S22 1.0 0.583465 0.583344 0.570792 Uiso ? S
S23 1.0 0.583237 0.333457 0.570766 Uiso ? S
S24 1.0 0.583236 0.083271 0.570766 Uiso ? S
S25 1.0 0.333764 0.833100 0.570729 Uiso ? S
S26 1.0 0.333579 0.583116 0.570769 Uiso ? S
S27 1.0 0.333401 0.333282 0.570800 Uiso ? S
S28 1.0 0.333579 0.083268 0.570768 Uiso ? S
S29 1.0 0.083525 0.833043 0.570672 Uiso ? S
S30 1.0 0.083396 0.583114 0.570770 Uiso ? S
S31 1.0 0.083397 0.333453 0.570768 Uiso ? S
S32 1.0 0.083525 0.083394 0.570670 Uiso ? S
B1 1.0 0.933387 0.933369 0.500684 Uiso ? B
B2 1.0 0.933422 0.733321 0.502035 Uiso ? B
B3 1.0 0.933415 0.533291 0.502801 Uiso ? B
B4 1.0 0.933421 0.333257 0.502013 Uiso ? B
B5 1.0 0.933386 0.133243 0.500664 Uiso ? B
B6 1.0 0.733338 0.933402 0.501992 Uiso ? B
B7 1.0 0.733407 0.733390 0.503161 Uiso ? B
B8 1.0 0.733407 0.533200 0.503161 Uiso ? B
B9 1.0 0.733337 0.333258 0.501988 Uiso ? B
B10 1.0 0.733381 0.133309 0.501179 Uiso ? B
B11 1.0 0.533309 0.933397 0.502768 Uiso ? B
B12 1.0 0.533219 0.733389 0.503160 Uiso ? B
B13 1.0 0.533310 0.533291 0.502781 Uiso ? B
B14 1.0 0.533378 0.333291 0.502451 Uiso ? B
B15 1.0 0.533379 0.133328 0.502446 Uiso ? B
B16 1.0 0.333273 0.933403 0.502014 Uiso ? B
B17 1.0 0.333272 0.733321 0.502033 Uiso ? B
B18 1.0 0.333310 0.533363 0.502493 Uiso ? B
B19 1.0 0.333346 0.333325 0.502644 Uiso ? B
B20 1.0 0.333309 0.133326 0.502468 Uiso ? B
B21 1.0 0.133257 0.933370 0.500702 Uiso ? B
B22 1.0 0.133324 0.733367 0.501264 Uiso ? B
B23 1.0 0.133343 0.533362 0.502503 Uiso ? B
B24 1.0 0.133345 0.333292 0.502483 Uiso ? B
B25 1.0 0.133324 0.133308 0.501223 Uiso ? B
N1 1.0 0.866712 0.866691 0.501224 Uiso ? N
N2 1.0 0.866699 0.666766 0.502929 Uiso ? N
N3 1.0 0.866699 0.466531 0.502923 Uiso ? N
N4 1.0 0.866711 0.266596 0.501207 Uiso ? N
N5 1.0 0.866754 0.066622 0.500492 Uiso ? N
N6 1.0 0.666783 0.866682 0.502898 Uiso ? N
N7 1.0 0.666678 0.666660 0.503388 Uiso ? N
N8 1.0 0.666784 0.466532 0.502903 Uiso ? N
N9 1.0 0.666730 0.266645 0.501794 Uiso ? N
N10 1.0 0.666731 0.066623 0.501797 Uiso ? N
N11 1.0 0.466551 0.866681 0.502907 Uiso ? N
N12 1.0 0.466551 0.666765 0.502920 Uiso ? N
N13 1.0 0.466641 0.466621 0.502731 Uiso ? N
N14 1.0 0.466795 0.266601 0.502654 Uiso ? N

N15 1.0 0.466641 0.066735 0.502715 Uiso ? N
N16 1.0 0.266611 0.866693 0.501238 Uiso ? N
N17 1.0 0.266663 0.666713 0.501878 Uiso ? N
N18 1.0 0.266619 0.466777 0.502690 Uiso ? N
N19 1.0 0.266620 0.266600 0.502673 Uiso ? N
N20 1.0 0.266662 0.066621 0.501839 Uiso ? N
N21 1.0 0.066633 0.866747 0.500575 Uiso ? N
N22 1.0 0.066639 0.666713 0.501880 Uiso ? N
N23 1.0 0.066754 0.466621 0.502754 Uiso ? N
N24 1.0 0.066639 0.266646 0.501842 Uiso ? N
N25 1.0 0.066636 0.066620 0.500532 Uiso ? N
F1 1.0 0.927170 0.536248 0.447765 Uiso ? F
F2 1.0 0.930581 0.534034 0.365511 Uiso ? F
F3 1.0 0.731113 0.134052 0.444017 Uiso ? F
F4 1.0 0.742378 0.128576 0.361828 Uiso ? F
F5 1.0 0.536111 0.927394 0.447657 Uiso ? F
F6 1.0 0.534438 0.931111 0.365410 Uiso ? F
F7 1.0 0.133270 0.734097 0.444810 Uiso ? F
F8 1.0 0.128441 0.743349 0.362602 Uiso ? F
F9 1.0 0.535422 0.534044 0.365429 Uiso ? F
F10 1.0 0.536038 0.536489 0.447678 Uiso ? F
F11 1.0 0.332206 0.334575 0.447677 Uiso ? F
F12 1.0 0.134232 0.133764 0.444417 Uiso ? F
F13 1.0 0.128419 0.128521 0.362226 Uiso ? F
F14 1.0 0.334044 0.332453 0.365435 Uiso ? F
Xe1 1.0 0.928031 0.535565 0.406591 Uiso ? Xe
Xe2 1.0 0.737578 0.130899 0.402805 Uiso ? Xe
Xe3 1.0 0.535700 0.928371 0.406485 Uiso ? Xe
Xe4 1.0 0.130507 0.739432 0.403603 Uiso ? Xe
Xe5 1.0 0.536190 0.535679 0.406506 Uiso ? Xe
Xe6 1.0 0.333140 0.333500 0.406509 Uiso ? Xe
Xe7 1.0 0.130943 0.130761 0.403210 Uiso ? Xe

## $WSe_2$ | hBN | $KrF_2$ after CT

_cell_length_a 22.616266
_cell_length_b 22.616266
_cell_length_c 38.172848
_cell_angle_alpha 90.000000
_cell_angle_beta 90.000000
_cell_angle_gamma 60.000000
_cell_volume 16909.353692
_space_group_name_H-M_alt 'P 1'
_space_group_IT_number 1

loop_
_space_group_symop_operation_xyz
'x, y, z'

```
loop_
  _atom_site_label
  _atom_site_occupancy
  _atom_site_fract_x
  _atom_site_fract_y
  _atom_site_fract_z
  _atom_site_adp_type
  _atom_site_U_iso_or_equiv
  _atom_site_type_symbol
  W1      1.0   0.952384    0.952384    0.639062   Uiso  ? W
  W2      1.0   0.809527    0.952374    0.639045   Uiso  ? W
  W3      1.0   0.952374    0.809527    0.639045   Uiso  ? W
  W4      1.0   0.666673    0.952369    0.639046   Uiso  ? W
  W5      1.0   0.809522    0.809522    0.639057   Uiso  ? W
  W6      1.0   0.952369    0.666673    0.639046   Uiso  ? W
  W7      1.0   0.523815    0.952374    0.639051   Uiso  ? W
  W8      1.0   0.666667    0.809535    0.638991   Uiso  ? W
  W9      1.0   0.809535    0.666667    0.638991   Uiso  ? W
  W10     1.0   0.952374    0.523815    0.639051   Uiso  ? W
  W11     1.0   0.380962    0.952364    0.639039   Uiso  ? W
  W12     1.0   0.523804    0.809530    0.638981   Uiso  ? W
  W13     1.0   0.666672    0.666671    0.638911   Uiso  ? W
  W14     1.0   0.809530    0.523804    0.638981   Uiso  ? W
  W15     1.0   0.952364    0.380961    0.639039   Uiso  ? W
  W16     1.0   0.238098    0.952376    0.639043   Uiso  ? W
  W17     1.0   0.380961    0.809523    0.639028   Uiso  ? W
  W18     1.0   0.523792    0.666681    0.638958   Uiso  ? W
  W19     1.0   0.666681    0.523792    0.638958   Uiso  ? W
  W20     1.0   0.809523    0.380961    0.639028   Uiso  ? W
  W21     1.0   0.952376    0.238098    0.639043   Uiso  ? W
  W22     1.0   0.095246    0.952380    0.639062   Uiso  ? W
  W23     1.0   0.238103    0.809527    0.639033   Uiso  ? W
  W24     1.0   0.380966    0.666671    0.639011   Uiso  ? W
  W25     1.0   0.523819    0.523819    0.639007   Uiso  ? W
  W26     1.0   0.666671    0.380966    0.639011   Uiso  ? W
  W27     1.0   0.809527    0.238103    0.639033   Uiso  ? W
  W28     1.0   0.952380    0.095246    0.639062   Uiso  ? W
  W29     1.0   0.095240    0.809525    0.639016   Uiso  ? W
  W30     1.0   0.238102    0.666671    0.638977   Uiso  ? W
  W31     1.0   0.380965    0.523805    0.639020   Uiso  ? W
  W32     1.0   0.523805    0.380965    0.639020   Uiso  ? W
  W33     1.0   0.666671    0.238101    0.638977   Uiso  ? W
  W34     1.0   0.809525    0.095240    0.639016   Uiso  ? W
  W35     1.0   0.095240    0.666667    0.638988   Uiso  ? W
  W36     1.0   0.238100    0.523811    0.639011   Uiso  ? W
  W37     1.0   0.380962    0.380962    0.639055   Uiso  ? W
  W38     1.0   0.523811    0.238100    0.639011   Uiso  ? W
  W39     1.0   0.666667    0.095240    0.638988   Uiso  ? W
  W40     1.0   0.095237    0.523807    0.639042   Uiso  ? W
  W41     1.0   0.238093    0.380961    0.639057   Uiso  ? W
  W42     1.0   0.380961    0.238092    0.639057   Uiso  ? W
  W43     1.0   0.523807    0.095237    0.639042   Uiso  ? W
  W44     1.0   0.095239    0.380966    0.639042   Uiso  ? W
  W45     1.0   0.238099    0.238099    0.639017   Uiso  ? W
  W46     1.0   0.380966    0.095239    0.639042   Uiso  ? W
  W47     1.0   0.095235    0.238100    0.638986   Uiso  ? W
  W48     1.0   0.238101    0.095235    0.638986   Uiso  ? W
  W49     1.0   0.095240    0.095240    0.639016   Uiso  ? W
  Se1     1.0   0.904766    0.904765    0.594317   Uiso  ? Se
  Se2     1.0   0.761922    0.904736    0.594322   Uiso  ? Se
  Se3     1.0   0.904737    0.761922    0.594322   Uiso  ? Se
  Se4     1.0   0.619021    0.904826    0.594296   Uiso  ? Se
  Se5     1.0   0.761950    0.761950    0.594284   Uiso  ? Se
  Se6     1.0   0.904827    0.619021    0.594296   Uiso  ? Se
```

Se7 1.0 0.476157 0.904830 0.594295 Uiso ? Se
Se8 1.0 0.619014 0.762001 0.594212 Uiso ? Se
Se9 1.0 0.762001 0.619013 0.594212 Uiso ? Se
Se10 1.0 0.904831 0.476157 0.594295 Uiso ? Se
Se11 1.0 0.333323 0.904763 0.594308 Uiso ? Se
Se12 1.0 0.476110 0.761969 0.594250 Uiso ? Se
Se13 1.0 0.619020 0.619020 0.594190 Uiso ? Se
Se14 1.0 0.761969 0.476110 0.594250 Uiso ? Se
Se15 1.0 0.904763 0.333323 0.594308 Uiso ? Se
Se16 1.0 0.190471 0.904775 0.594313 Uiso ? Se
Se17 1.0 0.333330 0.761937 0.594293 Uiso ? Se
Se18 1.0 0.476158 0.619038 0.594256 Uiso ? Se
Se19 1.0 0.619038 0.476158 0.594256 Uiso ? Se
Se20 1.0 0.761937 0.333330 0.594293 Uiso ? Se
Se21 1.0 0.904775 0.190471 0.594313 Uiso ? Se
Se22 1.0 0.047592 0.904823 0.594326 Uiso ? Se
Se23 1.0 0.190482 0.761968 0.594269 Uiso ? Se
Se24 1.0 0.333393 0.619020 0.594273 Uiso ? Se
Se25 1.0 0.476190 0.476189 0.594289 Uiso ? Se
Se26 1.0 0.619020 0.333393 0.594273 Uiso ? Se
Se27 1.0 0.761968 0.190481 0.594269 Uiso ? Se
Se28 1.0 0.904823 0.047592 0.594326 Uiso ? Se
Se29 1.0 0.047550 0.761958 0.594279 Uiso ? Se
Se30 1.0 0.190492 0.619008 0.594241 Uiso ? Se
Se31 1.0 0.333368 0.476137 0.594295 Uiso ? Se
Se32 1.0 0.476137 0.333368 0.594295 Uiso ? Se
Se33 1.0 0.619008 0.190492 0.594241 Uiso ? Se
Se34 1.0 0.761958 0.047550 0.594279 Uiso ? Se
Se35 1.0 0.047575 0.619023 0.594299 Uiso ? Se
Se36 1.0 0.190480 0.476145 0.594302 Uiso ? Se
Se37 1.0 0.333336 0.333336 0.594339 Uiso ? Se
Se38 1.0 0.476145 0.190480 0.594302 Uiso ? Se
Se39 1.0 0.619023 0.047575 0.594299 Uiso ? Se
Se40 1.0 0.047612 0.476198 0.594326 Uiso ? Se
Se41 1.0 0.190490 0.333374 0.594306 Uiso ? Se
Se42 1.0 0.333374 0.190490 0.594306 Uiso ? Se
Se43 1.0 0.476198 0.047612 0.594326 Uiso ? Se
Se44 1.0 0.047580 0.333409 0.594293 Uiso ? Se
Se45 1.0 0.190499 0.190499 0.594247 Uiso ? Se
Se46 1.0 0.333409 0.047580 0.594293 Uiso ? Se
Se47 1.0 0.047556 0.190491 0.594276 Uiso ? Se
Se48 1.0 0.190491 0.047556 0.594276 Uiso ? Se
Se49 1.0 0.047593 0.047593 0.594321 Uiso ? Se
Se50 1.0 0.904763 0.904762 0.683797 Uiso ? Se
Se51 1.0 0.761905 0.904765 0.683799 Uiso ? Se
Se52 1.0 0.904766 0.761904 0.683799 Uiso ? Se
Se53 1.0 0.619067 0.904735 0.683772 Uiso ? Se
Se54 1.0 0.761885 0.761885 0.683755 Uiso ? Se
Se55 1.0 0.904735 0.619067 0.683772 Uiso ? Se
Se56 1.0 0.476205 0.904728 0.683764 Uiso ? Se
Se57 1.0 0.619068 0.761870 0.683678 Uiso ? Se
Se58 1.0 0.761870 0.619068 0.683678 Uiso ? Se
Se59 1.0 0.904728 0.476205 0.683764 Uiso ? Se
Se60 1.0 0.333334 0.904760 0.683783 Uiso ? Se
Se61 1.0 0.476231 0.761883 0.683724 Uiso ? Se
Se62 1.0 0.619070 0.619070 0.683640 Uiso ? Se
Se63 1.0 0.761883 0.476231 0.683724 Uiso ? Se
Se64 1.0 0.904760 0.333334 0.683783 Uiso ? Se
Se65 1.0 0.190483 0.904759 0.683793 Uiso ? Se
Se66 1.0 0.333347 0.761893 0.683762 Uiso ? Se
Se67 1.0 0.476206 0.619064 0.683723 Uiso ? Se
Se68 1.0 0.619064 0.476206 0.683723 Uiso ? Se
Se69 1.0 0.761894 0.333347 0.683762 Uiso ? Se
Se70 1.0 0.904759 0.190483 0.683793 Uiso ? Se

Se71 1.0 0.047633 0.904739 0.683789 Uiso ? Se
Se72 1.0 0.190483 0.761883 0.683741 Uiso ? Se
Se73 1.0 0.333310 0.619067 0.683725 Uiso ? Se
Se74 1.0 0.476201 0.476201 0.683752 Uiso ? Se
Se75 1.0 0.619067 0.333310 0.683725 Uiso ? Se
Se76 1.0 0.761883 0.190482 0.683741 Uiso ? Se
Se77 1.0 0.904739 0.047633 0.683789 Uiso ? Se
Se78 1.0 0.047640 0.761881 0.683751 Uiso ? Se
Se79 1.0 0.190470 0.619069 0.683716 Uiso ? Se
Se80 1.0 0.333318 0.476210 0.683774 Uiso ? Se
Se81 1.0 0.476210 0.333318 0.683774 Uiso ? Se
Se82 1.0 0.619070 0.190469 0.683716 Uiso ? Se
Se83 1.0 0.761881 0.047639 0.683751 Uiso ? Se
Se84 1.0 0.047639 0.619063 0.683757 Uiso ? Se
Se85 1.0 0.190483 0.476208 0.683783 Uiso ? Se
Se86 1.0 0.333337 0.333337 0.683798 Uiso ? Se
Se87 1.0 0.476209 0.190483 0.683783 Uiso ? Se
Se88 1.0 0.619063 0.047639 0.683757 Uiso ? Se
Se89 1.0 0.047625 0.476191 0.683796 Uiso ? Se
Se90 1.0 0.190485 0.333316 0.683786 Uiso ? Se
Se91 1.0 0.333316 0.190485 0.683786 Uiso ? Se
Se92 1.0 0.476191 0.047624 0.683796 Uiso ? Se
Se93 1.0 0.047639 0.333306 0.683750 Uiso ? Se
Se94 1.0 0.190469 0.190469 0.683722 Uiso ? Se
Se95 1.0 0.333306 0.047639 0.683750 Uiso ? Se
Se96 1.0 0.047645 0.190481 0.683750 Uiso ? Se
Se97 1.0 0.190481 0.047645 0.683750 Uiso ? Se
Se98 1.0 0.047635 0.047635 0.683788 Uiso ? Se
B1 1.0 0.925923 0.925924 0.500056 Uiso ? B
B2 1.0 0.814772 0.926073 0.500245 Uiso ? B
B3 1.0 0.926072 0.814772 0.500245 Uiso ? B
B4 1.0 0.703664 0.925988 0.500221 Uiso ? B
B5 1.0 0.814781 0.814781 0.500511 Uiso ? B
B6 1.0 0.925988 0.703665 0.500221 Uiso ? B
B7 1.0 0.592525 0.926005 0.500075 Uiso ? B
B8 1.0 0.703705 0.814757 0.499643 Uiso ? B
B9 1.0 0.814756 0.703705 0.499643 Uiso ? B
B10 1.0 0.926004 0.592525 0.500075 Uiso ? B
B11 1.0 0.481517 0.925936 0.500290 Uiso ? B
B12 1.0 0.592628 0.814749 0.498856 Uiso ? B
B13 1.0 0.703762 0.703762 0.497541 Uiso ? B
B14 1.0 0.814748 0.592629 0.498856 Uiso ? B
B15 1.0 0.925937 0.481517 0.500290 Uiso ? B
B16 1.0 0.370330 0.925941 0.500309 Uiso ? B
B17 1.0 0.481555 0.814680 0.499421 Uiso ? B
B18 1.0 0.592418 0.703913 0.497576 Uiso ? B
B19 1.0 0.703913 0.592418 0.497576 Uiso ? B
B20 1.0 0.814680 0.481556 0.499421 Uiso ? B
B21 1.0 0.925940 0.370330 0.500309 Uiso ? B
B22 1.0 0.259251 0.926034 0.500111 Uiso ? B
B23 1.0 0.370402 0.814762 0.500789 Uiso ? B
B24 1.0 0.481487 0.703871 0.499421 Uiso ? B
B25 1.0 0.592591 0.592591 0.498379 Uiso ? B
B26 1.0 0.703870 0.481488 0.499421 Uiso ? B
B27 1.0 0.814762 0.370403 0.500789 Uiso ? B
B28 1.0 0.926033 0.259252 0.500110 Uiso ? B
B29 1.0 0.148149 0.925973 0.500219 Uiso ? B
B30 1.0 0.259291 0.814750 0.500343 Uiso ? B
B31 1.0 0.370256 0.703750 0.500080 Uiso ? B
B32 1.0 0.481462 0.592436 0.500042 Uiso ? B
B33 1.0 0.592436 0.481462 0.500042 Uiso ? B
B34 1.0 0.703749 0.370256 0.500080 Uiso ? B
B35 1.0 0.814750 0.259292 0.500343 Uiso ? B
B36 1.0 0.925973 0.148149 0.500219 Uiso ? B

B37 1.0 0.036967 0.926013 0.500249 Uiso ? B
B38 1.0 0.148217 0.814780 0.498812 Uiso ? B
B39 1.0 0.259297 0.703690 0.498675 Uiso ? B
B40 1.0 0.370349 0.592613 0.499720 Uiso ? B
B41 1.0 0.481515 0.481515 0.500260 Uiso ? B
B42 1.0 0.592613 0.370349 0.499720 Uiso ? B
B43 1.0 0.703690 0.259297 0.498675 Uiso ? B
B44 1.0 0.814780 0.148217 0.498812 Uiso ? B
B45 1.0 0.926013 0.036967 0.500249 Uiso ? B
B46 1.0 0.036972 0.814776 0.499016 Uiso ? B
B47 1.0 0.148155 0.703643 0.497351 Uiso ? B
B48 1.0 0.259249 0.592508 0.497837 Uiso ? B
B49 1.0 0.370279 0.481530 0.499956 Uiso ? B
B50 1.0 0.481530 0.370279 0.499956 Uiso ? B
B51 1.0 0.592507 0.259249 0.497837 Uiso ? B
B52 1.0 0.703642 0.148155 0.497351 Uiso ? B
B53 1.0 0.814775 0.036972 0.499016 Uiso ? B
B54 1.0 0.037039 0.703731 0.498400 Uiso ? B
B55 1.0 0.148199 0.592586 0.497999 Uiso ? B
B56 1.0 0.259218 0.481566 0.499456 Uiso ? B
B57 1.0 0.370379 0.370379 0.500392 Uiso ? B
B58 1.0 0.481566 0.259218 0.499456 Uiso ? B
B59 1.0 0.592586 0.148199 0.497999 Uiso ? B
B60 1.0 0.703731 0.037040 0.498400 Uiso ? B
B61 1.0 0.037036 0.592573 0.499642 Uiso ? B
B62 1.0 0.148139 0.481530 0.499708 Uiso ? B
B63 1.0 0.259287 0.370242 0.500572 Uiso ? B
B64 1.0 0.370242 0.259287 0.500572 Uiso ? B
B65 1.0 0.481529 0.148139 0.499708 Uiso ? B
B66 1.0 0.592573 0.037037 0.499642 Uiso ? B
B67 1.0 0.036999 0.481560 0.500437 Uiso ? B
B68 1.0 0.148251 0.370228 0.499881 Uiso ? B
B69 1.0 0.259268 0.259269 0.499281 Uiso ? B
B70 1.0 0.370228 0.148251 0.499882 Uiso ? B
B71 1.0 0.481560 0.036999 0.500437 Uiso ? B
B72 1.0 0.037019 0.370378 0.499447 Uiso ? B
B73 1.0 0.148239 0.259293 0.497996 Uiso ? B
B74 1.0 0.259292 0.148239 0.497996 Uiso ? B
B75 1.0 0.370378 0.037020 0.499447 Uiso ? B
B76 1.0 0.036993 0.259217 0.498562 Uiso ? B
B77 1.0 0.148130 0.148130 0.497153 Uiso ? B
B78 1.0 0.259216 0.036993 0.498562 Uiso ? B
B79 1.0 0.037001 0.148218 0.499010 Uiso ? B
B80 1.0 0.148218 0.037002 0.499010 Uiso ? B
B81 1.0 0.037035 0.037035 0.500071 Uiso ? B
N1 1.0 0.963030 0.963030 0.500643 Uiso ? N
N2 1.0 0.851694 0.963063 0.500140 Uiso ? N
N3 1.0 0.963063 0.851694 0.500140 Uiso ? N
N4 1.0 0.740786 0.963011 0.499974 Uiso ? N
N5 1.0 0.851874 0.851874 0.501034 Uiso ? N
N6 1.0 0.963011 0.740786 0.499974 Uiso ? N
N7 1.0 0.629654 0.963000 0.500391 Uiso ? N
N8 1.0 0.740736 0.851795 0.500664 Uiso ? N
N9 1.0 0.851795 0.740736 0.500664 Uiso ? N
N10 1.0 0.963000 0.629653 0.500391 Uiso ? N
N11 1.0 0.518489 0.962977 0.500706 Uiso ? N
N12 1.0 0.629679 0.851773 0.499678 Uiso ? N
N13 1.0 0.740770 0.740770 0.498923 Uiso ? N
N14 1.0 0.851773 0.629679 0.499678 Uiso ? N
N15 1.0 0.962978 0.518488 0.500706 Uiso ? N
N16 1.0 0.407389 0.962995 0.500378 Uiso ? N
N17 1.0 0.518496 0.851886 0.499670 Uiso ? N
N18 1.0 0.629527 0.740790 0.497456 Uiso ? N
N19 1.0 0.740789 0.629527 0.497456 Uiso ? N

N20 1.0 0.851886 0.518496 0.499670 Uiso ? N
N21 1.0 0.962996 0.407389 0.500378 Uiso ? N
N22 1.0 0.296332 0.963007 0.499967 Uiso ? N
N23 1.0 0.407415 0.851893 0.500661 Uiso ? N
N24 1.0 0.518543 0.740668 0.498524 Uiso ? N
N25 1.0 0.629669 0.629668 0.497273 Uiso ? N
N26 1.0 0.740667 0.518543 0.498524 Uiso ? N
N27 1.0 0.851892 0.407415 0.500661 Uiso ? N
N28 1.0 0.963008 0.296331 0.499967 Uiso ? N
N29 1.0 0.185161 0.963005 0.500142 Uiso ? N
N30 1.0 0.296313 0.851787 0.501049 Uiso ? N
N31 1.0 0.407468 0.740751 0.500456 Uiso ? N
N32 1.0 0.518439 0.629661 0.499257 Uiso ? N
N33 1.0 0.629662 0.518439 0.499256 Uiso ? N
N34 1.0 0.740751 0.407468 0.500456 Uiso ? N
N35 1.0 0.851787 0.296313 0.501050 Uiso ? N
N36 1.0 0.963005 0.185161 0.500142 Uiso ? N
N37 1.0 0.074001 0.962974 0.500643 Uiso ? N
N38 1.0 0.185241 0.851779 0.500138 Uiso ? N
N39 1.0 0.296202 0.740760 0.500003 Uiso ? N
N40 1.0 0.407448 0.629543 0.500166 Uiso ? N
N41 1.0 0.518550 0.518550 0.500281 Uiso ? N
N42 1.0 0.629543 0.407448 0.500166 Uiso ? N
N43 1.0 0.740760 0.296202 0.500002 Uiso ? N
N44 1.0 0.851778 0.185241 0.500138 Uiso ? N
N45 1.0 0.962974 0.074001 0.500643 Uiso ? N
N46 1.0 0.074055 0.851843 0.499519 Uiso ? N
N47 1.0 0.185261 0.740720 0.498015 Uiso ? N
N48 1.0 0.296307 0.629681 0.498599 Uiso ? N
N49 1.0 0.407362 0.518484 0.500278 Uiso ? N
N50 1.0 0.518484 0.407362 0.500278 Uiso ? N
N51 1.0 0.629681 0.296307 0.498599 Uiso ? N
N52 1.0 0.740720 0.185261 0.498015 Uiso ? N
N53 1.0 0.851844 0.074054 0.499519 Uiso ? N
N54 1.0 0.074000 0.740820 0.498023 Uiso ? N
N55 1.0 0.185154 0.629617 0.497291 Uiso ? N
N56 1.0 0.296313 0.518483 0.499125 Uiso ? N
N57 1.0 0.407390 0.407390 0.500538 Uiso ? N
N58 1.0 0.518483 0.296313 0.499125 Uiso ? N
N59 1.0 0.629617 0.185154 0.497291 Uiso ? N
N60 1.0 0.740820 0.074000 0.498023 Uiso ? N
N61 1.0 0.074060 0.629662 0.498594 Uiso ? N
N62 1.0 0.185183 0.518519 0.499135 Uiso ? N
N63 1.0 0.296263 0.407425 0.500746 Uiso ? N
N64 1.0 0.407425 0.296264 0.500746 Uiso ? N
N65 1.0 0.518520 0.185182 0.499135 Uiso ? N
N66 1.0 0.629662 0.074060 0.498594 Uiso ? N
N67 1.0 0.074110 0.518506 0.500304 Uiso ? N
N68 1.0 0.185195 0.407415 0.500549 Uiso ? N
N69 1.0 0.296281 0.296283 0.500738 Uiso ? N
N70 1.0 0.407414 0.185196 0.500549 Uiso ? N
N71 1.0 0.518506 0.074110 0.500304 Uiso ? N
N72 1.0 0.074126 0.407383 0.500301 Uiso ? N
N73 1.0 0.185190 0.296273 0.499131 Uiso ? N
N74 1.0 0.296273 0.185190 0.499131 Uiso ? N
N75 1.0 0.407383 0.074126 0.500301 Uiso ? N
N76 1.0 0.074123 0.296257 0.498586 Uiso ? N
N77 1.0 0.185194 0.185194 0.497286 Uiso ? N
N78 1.0 0.296257 0.074123 0.498587 Uiso ? N
N79 1.0 0.074094 0.185192 0.498009 Uiso ? N
N80 1.0 0.185192 0.074094 0.498009 Uiso ? N
N81 1.0 0.074072 0.074072 0.499512 Uiso ? N
Kr1 1.0 0.925859 0.925858 0.370081 Uiso ? Kr
Kr2 1.0 0.814892 0.814894 0.370120 Uiso ? Kr

Kr3 1.0 0.592562 0.925713 0.370010 Uiso ? Kr
Kr4 1.0 0.925713 0.592566 0.370010 Uiso ? Kr
Kr5 1.0 0.703573 0.703571 0.369951 Uiso ? Kr
Kr6 1.0 0.481747 0.814564 0.369906 Uiso ? Kr
Kr7 1.0 0.814569 0.481743 0.369906 Uiso ? Kr
Kr8 1.0 0.259460 0.925583 0.369984 Uiso ? Kr
Kr9 1.0 0.592454 0.592469 0.373451 Uiso ? Kr
Kr10 1.0 0.925583 0.259451 0.369982 Uiso ? Kr
Kr11 1.0 0.370763 0.703500 0.370030 Uiso ? Kr
Kr12 1.0 0.703499 0.370764 0.370031 Uiso ? Kr
Kr13 1.0 0.148482 0.814871 0.370030 Uiso ? Kr
Kr14 1.0 0.481746 0.481747 0.369901 Uiso ? Kr
Kr15 1.0 0.814872 0.148482 0.370030 Uiso ? Kr
Kr16 1.0 0.259501 0.592404 0.369856 Uiso ? Kr
Kr17 1.0 0.592410 0.259496 0.369856 Uiso ? Kr
Kr18 1.0 0.036918 0.703610 0.370102 Uiso ? Kr
Kr19 1.0 0.370379 0.370378 0.370168 Uiso ? Kr
Kr20 1.0 0.703614 0.036919 0.370101 Uiso ? Kr
Kr21 1.0 0.148270 0.481570 0.370124 Uiso ? Kr
Kr22 1.0 0.481570 0.148268 0.370124 Uiso ? Kr
Kr23 1.0 0.259210 0.259207 0.370029 Uiso ? Kr
Kr24 1.0 0.036740 0.370400 0.370130 Uiso ? Kr
Kr25 1.0 0.370395 0.036745 0.370131 Uiso ? Kr
Kr26 1.0 0.148236 0.148242 0.369964 Uiso ? Kr
Kr27 1.0 0.036955 0.036954 0.370185 Uiso ? Kr
F1 1.0 0.925800 0.925800 0.421591 Uiso ? F
F2 1.0 0.814982 0.814983 0.421626 Uiso ? F
F3 1.0 0.592698 0.925937 0.421435 Uiso ? F
F4 1.0 0.925938 0.592699 0.421435 Uiso ? F
F5 1.0 0.703526 0.703527 0.421315 Uiso ? F
F6 1.0 0.481671 0.814414 0.421291 Uiso ? F
F7 1.0 0.814416 0.481669 0.421290 Uiso ? F
F8 1.0 0.259200 0.925857 0.421384 Uiso ? F
F9 1.0 0.592516 0.592540 0.425070 Uiso ? F
F10 1.0 0.925860 0.259199 0.421379 Uiso ? F
F11 1.0 0.370784 0.703535 0.421448 Uiso ? F
F12 1.0 0.703536 0.370784 0.421448 Uiso ? F
F13 1.0 0.148250 0.814875 0.421519 Uiso ? F
F14 1.0 0.481701 0.481699 0.421301 Uiso ? F
F15 1.0 0.814877 0.148249 0.421519 Uiso ? F
F16 1.0 0.259099 0.592425 0.421272 Uiso ? F
F17 1.0 0.592427 0.259096 0.421272 Uiso ? F
F18 1.0 0.036866 0.703681 0.421657 Uiso ? F
F19 1.0 0.370210 0.370210 0.421667 Uiso ? F
F20 1.0 0.703678 0.036868 0.421657 Uiso ? F
F21 1.0 0.148346 0.481497 0.421678 Uiso ? F
F22 1.0 0.481496 0.148345 0.421679 Uiso ? F
F23 1.0 0.259201 0.259200 0.421526 Uiso ? F
F24 1.0 0.037002 0.370158 0.421671 Uiso ? F
F25 1.0 0.370161 0.036998 0.421670 Uiso ? F
F26 1.0 0.148258 0.148257 0.421505 Uiso ? F
F27 1.0 0.036962 0.036961 0.421775 Uiso ? F
F28 1.0 0.925897 0.925894 0.319134 Uiso ? F
F29 1.0 0.815012 0.815013 0.319174 Uiso ? F
F30 1.0 0.592589 0.926113 0.319226 Uiso ? F
F31 1.0 0.926114 0.592590 0.319226 Uiso ? F
F32 1.0 0.703679 0.703680 0.319048 Uiso ? F
F33 1.0 0.481680 0.814659 0.318988 Uiso ? F
F34 1.0 0.814664 0.481675 0.318989 Uiso ? F
F35 1.0 0.259107 0.925987 0.319211 Uiso ? F
F36 1.0 0.592406 0.592429 0.321950 Uiso ? F
F37 1.0 0.925989 0.259106 0.319206 Uiso ? F
F38 1.0 0.370788 0.703640 0.319109 Uiso ? F
F39 1.0 0.703641 0.370787 0.319110 Uiso ? F

F40 1.0 0.148137 0.814986 0.319133 Uiso ? F
F41 1.0 0.481592 0.481591 0.318989 Uiso ? F
F42 1.0 0.814987 0.148135 0.319133 Uiso ? F
F43 1.0 0.259132 0.592438 0.319098 Uiso ? F
F44 1.0 0.592441 0.259129 0.319098 Uiso ? F
F45 1.0 0.036949 0.703647 0.319125 Uiso ? F
F46 1.0 0.370276 0.370276 0.319250 Uiso ? F
F47 1.0 0.703648 0.036946 0.319125 Uiso ? F
F48 1.0 0.148232 0.481524 0.319124 Uiso ? F
F49 1.0 0.481524 0.148232 0.319123 Uiso ? F
F50 1.0 0.259192 0.259191 0.319108 Uiso ? F
F51 1.0 0.037028 0.370277 0.319198 Uiso ? F
F52 1.0 0.370276 0.037027 0.319198 Uiso ? F
F53 1.0 0.148261 0.148261 0.318990 Uiso ? F
F54 1.0 0.036945 0.036945 0.319169 Uiso ? F

## S2. Bond length changes upon CT and average geometric polarization of the BN separator layer under CT

**Table S1.** Comparison of the average bond lengths [Å] before and after charge transfer in the investigated $WS_2$ | hBN | $XeF_2$ and $WSe_2$ | hBN | $KrF_2$ heterostructures. Relative stretching for W–Se and Kr–F bonds reached 1% and 3.1%, respectively, whereas for W–S and Xe–F bonds it is negligible and 2.6%, respectively. These effects come from CT, e.g. the gradual population of the Ng–F antibonding LUMO orbital caused the Ng–F bond elongation, etc.

| | Before CT | After CT |
|---|---|---|
| R (W–S) | 2.395 | 2.395 |
| R (Xe–F) | 1.988 | 2.041 |
| R (W–Se) | 2.504 | 2.529 |
| R (Kr–F) | 1.897 | 1.955 |

**Table S2.** Average geometric polarization of the BN monolayers for the analysed heterostructures in the electric field originating from the CT. $Z_{frac}(E)$ presents averaged position of the atom E in the fractional coordinates over the whole monolayer. For the case of the larger CT, i.e. $WSe_2$ | hBN | $KrF_2$ the hBN layer – initially flat – becomes slightly puckered with the separation of the N and B atoms in the crystallographic $\boldsymbol{c}$ direction of ca. 0.008 Å.

| | $WS_2$ \| hBN \| $XeF_2$ | $WSe_2$ \| hBN \| $KrF_2$ |
|---|---|---|
| $Z_{frac}(B)$ before CT | 0.502923 | 0.479323 |
| $Z_{frac}(N)$ before CT | 0.502888 | 0.479263 |
| $Z_{frac}(B)$ after CT | 0.502124 | 0.499437 |
| $Z_{frac}(N)$ after CT | 0.502135 | 0.499614 |

## S3. Example of a TMD | hBN | RED system

**Table S3.** The valence band maximum (VBM), conduction band minimum (CBM) and band gaps in units of eV for a range of TMDs with lithium as an example of a strong reductor. The energy scales were aligned with respect to the He 1s peak. In comparison to CBM of presented TMDs, lithium shows a type III behavior with a broken gap, giving possibility of high electron doping.

| | hBN | $MoSe_2$ | $MoS_2$ | $WSe_2$ | $WS_2$ | Li |
|---|---|---|---|---|---|---|
| VBM | 9.86 | 10.37 | 9.76 | 10.8 | 9.98 | 12.35 |
| CBM | 14.14 | 11.83 | 11.43 | 12.09 | 11.82 | 12.35 |
| Band gap | 4.28 | 1.46 | 1.67 | 1.29 | 1.84 | — |

**Table S4.** Values of $E_{Li}^{VBM}$-$E_{TMD}^{CBM}$ calculated for the selected TMDs. As in the case of the TMD | hBN | OX system, where strong electron acceptor in a broken gap system could induce large hole injection into TMD monolayers, a TMD/Li is a broken gap system, likely leading to large interface CT and electron doping into TMD monolayers.

| | $MoSe_2$ | $MoS_2$ | $WSe_2$ | $WS_2$ |
|---|---|---|---|---|
| Li | 0.61 | 1.01 | 0.35 | 0.62 |